\documentclass[aps,prl,twocolumn,superscriptaddress,floatfix,showpacs,groupedaddress]{revtex4-1}
\usepackage{hyperref}
\usepackage{graphicx}
\usepackage{amsfonts}
\usepackage{amssymb}
\usepackage{amsmath}
\usepackage{txfonts}
\usepackage{lipsum}
\usepackage{color}
\usepackage{wasysym}

\DeclareMathOperator{\Tr}{Tr}
\DeclareMathOperator{\Det}{Det}
\DeclareMathOperator{\sgn}{sgn}

\begin{document}
\title{Laser-induced topological transitions in phosphorene with inversion symmetry}
\author{C. Dutreix, E. A. Stepanov, M. I. Katsnelson}
\affiliation{Radboud University, Institute for Molecules and Materials, Heyendaalseweg 135, 6525AJ Nijmegen, The Netherlands}

\begin{abstract}
Recent \textit{ab initio} calculations and experiments reported insulating-semimetallic phase transitions in multilayer phosphorene under a perpendicular dc field, pressure or doping, as a possible route to realize topological phases. In this work, we show that even a monolayer phosphorene may undergo Lifshitz transitions toward semimetallic and topological insulating phases, provided it is rapidly driven by in-plane time-periodic laser fields. Based on a four-orbital tight-binding description, we give an inversion-symmetry-based prescription in order to apprehend the topology of the photon-renormalized band structure, up to the second order in the high-frequency limit. Apart from the initial band insulating behavior, two additional phases are thus identified. A semimetallic phase with massless Dirac electrons may be induced by linear polarized fields, whereas elliptic polarized fields are likely to drive the material into an anomalous quantum Hall phase.
\end{abstract}

\maketitle

More than a century ago, Bridgman reported his attempt to change phosphorus from white to red, which actually resulted in the discovery of the most thermodynamically stable phosphorus allotrope at room temperature, namely, black phosphorus (BP) \cite{bridgman1914two}. This is, like graphite and transition metal dichalcogenides, a layered material in which one-atomic-thick films are bound to each other via weak van der Waals interactions, enabling the isolation of a few layers by mechanical exfoliation \cite{li2014black, liu2014phosphorene, xia2014rediscovering}. Low-energy electrons reveal a strongly anisotropic semiconductor with a direct energy gap and high charge carrier mobility, features that make the phosphorus allotrope particularly suitable for electronic applications, such as field-effect transistors \cite{qiao2014high,Rudenko:2014yg}. From a fundamental perspective, investigations of pressure-induced electronic and optical properties in bulk BP addressed, a few decades ago, the possibilities to realize structural phase transitions and to tune the semiconducting gap \cite{PhysRev.92.580,PMID:17757066,1963JAP....34.1853W}. Recent \textit{ab initio} calculations and experiments pointed out that gap tuning under pressure, dc field, or doping, could even result in an insulating-semimetallic Lifshitz transition in multilayer BP, thus opening a route toward the realization of symmetry-protected topological phases in this non-compound material \cite{PhysRevLett.112.176801,PhysRevLett.113.046804,fei2015topologically,Rudenko:2015kq,liu2015switching,kim2015observation,baik2015emergence}.

Contrary to the above works, we focus here on monolayer BP, namely phosphorene, when it is rapidly driven by in-plane homogeneous time-periodic laser fields. Within a four-orbital tight-binding approximation introduced recently \cite{Rudenko:2015kq}, we provide an inversion-symmetry-based prescription that enables us to apprehend the band structure topology up to the second order in the high-frequency limit. Three distinct phases are then identified, depending on the field polarization and magnitude. In particular, linear polarized fields are likely to drive phosphorene from a band insulating phase toward a semimetallic one with two nonequivalent Dirac cones, thus mimicking electronic properties of graphene massless Dirac quasiparticles \cite{RevModPhys.81.109,katsnelson2012graphene}. Right at the transition, the universal semi-Dirac band structure, which describes massless electrons only along one direction, is recovered \cite{montambaux2009universal, dietl2008new,PhysRevB.87.245413}. Being the underlying mechanism of topological transitions \cite{Rudenko:2015kq,liu2015switching,kim2015observation,baik2015emergence}, this Lifshitz transition shows that non-trivial phases may be induced in phosphorene too. Besides, we show that, under elliptic polarizations, it may also undergo a particle-hole-symmetry-protected topological transition toward an anomalous quantum Hall phase, in which a non-zero Chern number characterizes the existence of chiral boundary modes. Therefore, these three distinct phases are predicted in monolayer BP by simply varying the polarization and intensity of a single external field, which is more controllable than pressure or doping, and is also suitable for realizations in ultracold atomic gases and photonic crystals \cite{tarruell2012creating,rechtsman2013photonic,PhysRevA.89.063628,PhysRevA.89.061603,jotzu2014experimental}.\\

\begin{figure}[b!]
\includegraphics[trim = 5mm 5mm 10mm 10mm, clip, width=5.cm]{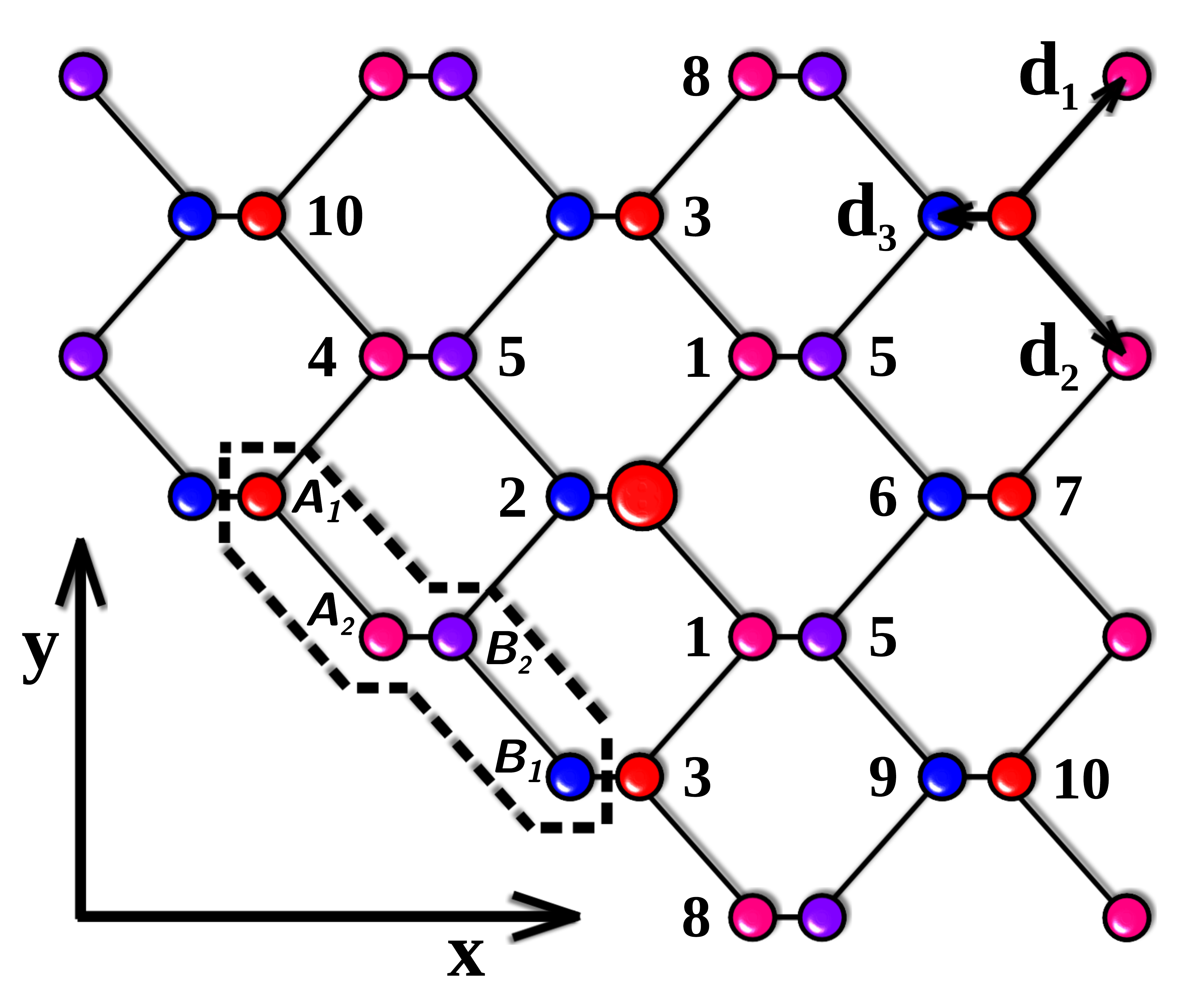}
\caption{\small (Color online) Top view of phosphorene. The black dashed line outlines the unit cell made of atoms $\textcolor[RGB]{255,0,0}{\CIRCLE}~\mbox{---}~$A$_1$, $\textcolor[RGB]{255,0,128}{\CIRCLE}~\mbox{---}~$A$_2$, $\textcolor[RGB]{128,0,255}{\CIRCLE}~\mbox{---}~$B$_2$, $\textcolor[RGB]{0,0,255}{\CIRCLE}~\mbox{---}~$B$_1$. Nearest-neighbor vectors are ${\bf d_{1}} = a_{0} (\cos \theta, \sin \theta)$, ${\bf d_{2}} = a_{0} (\cos \theta, -\sin \theta)$, and ${\bf d_{3}} =~ b_{0} (−1,0)$, with $\theta \simeq 48.395^{\circ}$, $a_{0}\simeq2.216~ \AA$ and $b_{0}\simeq0.716~ \AA$ \cite{Rudenko:2014yg}. Numbers from 1 to 10 refer to the ten different hoppings allowed from the central bigger $A_{1}$ site.}
\label{Lattice}
\end{figure}

{\it{Tight-binding description}} --- 
Phosphorene consists of a rectangular Bravais lattice and an out-of-plane unit cell made of four phosphorus atoms, each of which have three nearest neighbors, as illustrated in Fig. \ref{Lattice}.
Electronic properties are reasonably described within a tight-binding description that involves up to ten hopping processes \cite{Rudenko:2015kq}. The corresponding Bloch band structure for spinless electrons is then given by a $4\times4$ matrix which, in the sublattice basis $\{A_{1}, B_{2}, A_{2}, B_{1}\}$, can generically be written as
\begin{align}\label{Bloch Hamiltonian Matrix}
H({\bf k}) = 
\left( \begin{array}{llll} 
F_{3}({\bf k}) & F_{4}({\bf k}) & F_{1}({\bf k}) & F_{2}({\bf k}) \\
F_{4}^{*}({\bf k}) & F_{3}({\bf k}) & F_{2}({\bf k})~ e^{i\varphi({\bf k})} & F_{1}({\bf k}) \\
F_{1}^{*}({\bf k}) & F_{2}^{*}({\bf k})~ e^{-i\varphi({\bf k})} & F_{3}({\bf k}) & F_{4}({\bf k}) \\
F_{2}^{*}({\bf k}) & F_{1}^{*}({\bf k}) & F_{4}^{*}({\bf k}) & F_{3}({\bf k})
\end{array} \right) ~.
\end{align}
Matrix components $F_{i=1...4}$ are functions of hopping amplitudes and wavevector ${\bf k}$ such that $H({\bf k}+{\bf G})=H({\bf k})$, with ${\bf G}$ a reciprocal lattice vector \cite{SM}. Besides $\varphi({\bf k}) = k_{x}-k_{y}$ for $k_{x}$ and $k_{y}$ given in units of $a^{-1}_1$ and $a^{-1}_2$, respectively.\\

{\it{Time-periodic laser fields}} ---
Suppose that an in-plane homogeneous time-periodic electric field of frequency $\Omega$ is turned on, driving the system into a nonequilibrium steady state. How electrons relax into such a stabilized regime is currently under investigations \cite{PhysRevX.5.041050,PhysRevE.93.012130,PhysRevLett.116.120401}. The field yields a vector potential of the form ${\bf A}(t)=~( A_{x} \cos \Omega t, A_{y} \sin[\Omega t - \phi], 0 )$, whereas the scalar potential is wiped out, for we consider the temporal gauge. Hereinafter, it is assumed that $c=\hbar=1$, so that $A_{x\,(y)}=eE_{x\,(y)}\,\Omega^{-1}$ with $E_{x\,(y)}$ the $x\,(y)$-component of the field magnitude. The field polarization is elliptic for $\phi=0$ and linear for $\phi=\pi/2$. The vector potential subsequently enters matrix \eqref{Bloch Hamiltonian Matrix} via Peierls phases \cite{SM}.

The time evolution of a quantum state is ruled by the time-dependent Schr\"odinger equation $i \partial_\tau \Psi(\lambda, \tau) =~\lambda H(\tau) \Psi (\lambda, \tau)$,
where $\tau=\Omega t$ and frequency is encoded in the dimensionless parameter $\lambda=\delta E/ \Omega$. Energy scale $\delta E$ corresponds to phosphorene bandwidth ($14~eV$ \cite{Rudenko:2015kq}) which involves the four photon-renormalized energy bands. Besides, $H(\tau)$ is given in units of $\delta E$ and is dimensionless too. Here we are interested in the high-frequency limit $\lambda \ll 1$ and use an extension of Magnus expansion \cite{Magnus}, in the spirit of Ref. \cite{Itin2}. It relies on a unitary transformation defined by $\Psi (\lambda, \tau) = \exp\{-i\Delta(\tau)\}\,\psi (\lambda, \tau)$, where $\Delta(\tau) =~ \sum_{n=1}^{+\infty} \lambda^{n} \Delta_{n}(\tau)$, and where operators $\Delta_{n}(\tau)$ remove the $\tau$-dependent terms of $H(\tau)$. It is also assumed that $\Delta_{n}(\tau)$ has the same time periodicity as $H(\tau)$ and averages at zero \cite{PhysRevLett.115.075301}. This transformation leads to $i\partial_\tau \psi(\lambda, \tau) = \tilde H \psi (\lambda, \tau)$, with effective Hamiltonian
\begin{align}\label{Effective Hamiltonian}
\tilde{H}=\lambda e^{i\Delta(t)} H(t) e^{-i\Delta(t)} -i e^{i\Delta(t)} \partial_{t} e^{-i\Delta(t)} ~.
\end{align}
Snider's identity \cite{snider1964perturbation} implies
\begin{align}\label{Exponential Operator Derivative}
\partial_{\tau} e^{-i\Delta(\tau)} &= \sum_{n=0}^{\infty} \frac{ \big\{ \big(\!-i\Delta(\tau) \big)^{n}, -\,i \partial_{\tau}\Delta(\tau) \big\}}{(n+1)!}~ e^{-i\Delta(\tau)} ~,
\end{align}
where the repeated commutator is defined for two operators $X$ and $Y$ by $\{1, Y\}=Y$ and $\{X^{n}, Y\}=~[X, \{ X^{n-1}, Y\}]$  \cite{wilcox1967exponential}; the square brackets denote the usual commutator. Using the series representation $\tilde{H}=\sum_{n=1}^{\infty}\lambda^{n}\tilde{H}_{n}$, together with Eqs. (\ref{Effective Hamiltonian}) and (\ref{Exponential Operator Derivative}), one can then determine operators $\tilde H_{n}$ and $\Delta_{n}$ iteratively. Here, we restrict the high-frequency analysis to the second order in $\lambda$, which leads to
\begin{align}
\tilde H_{1} &= H(\tau)-\partial_{\tau}\Delta_{1}(\tau) ~, \\
\tilde H_{2} &= i\,[\Delta_{1}(\tau),H(\tau)]-\frac{i}{2}[\Delta_{1}(\tau),\partial_{t}\Delta_{1}(\tau)]-\partial_{\tau}\Delta_{2}(\tau) ~. \notag
\end{align}
Imposing these two operators to be static requires us to retain only $\tau$-independent terms in $\tilde H_{1}$ and $\tilde H_{2}$, which is achieved by time averaging over a period because $H(\tau)$ and $\Delta_{n}(\tau)$ are time periodic. So this averaging method leads to effective time-independent Hamiltonians that describe the stroboscopic dynamics, whereas the system evolution between two stroboscopic times is encoded in $\Delta_{n}(\tau)$. This finally results in
\begin{align}\label{Time-independent Hamiltonians}
\tilde H_{1} = H_{0} \text{~~~and~~~}
\tilde H_{2} = -\sum_{m>0}\frac{[H_{m},H_{-m}]}{m} ~,
\end{align}
where $H_{m}=\int_{-\pi}^{+\pi} \frac{d\tau}{2\pi}~ e^{im\tau} H(\tau)$ defines the Floquet matrix form of the Hamiltonian.
\begin{figure}[!t]
\centering
$\begin{array}{cc}
\includegraphics[trim = 0mm 0mm 0mm 0mm, clip, width=3.6cm]{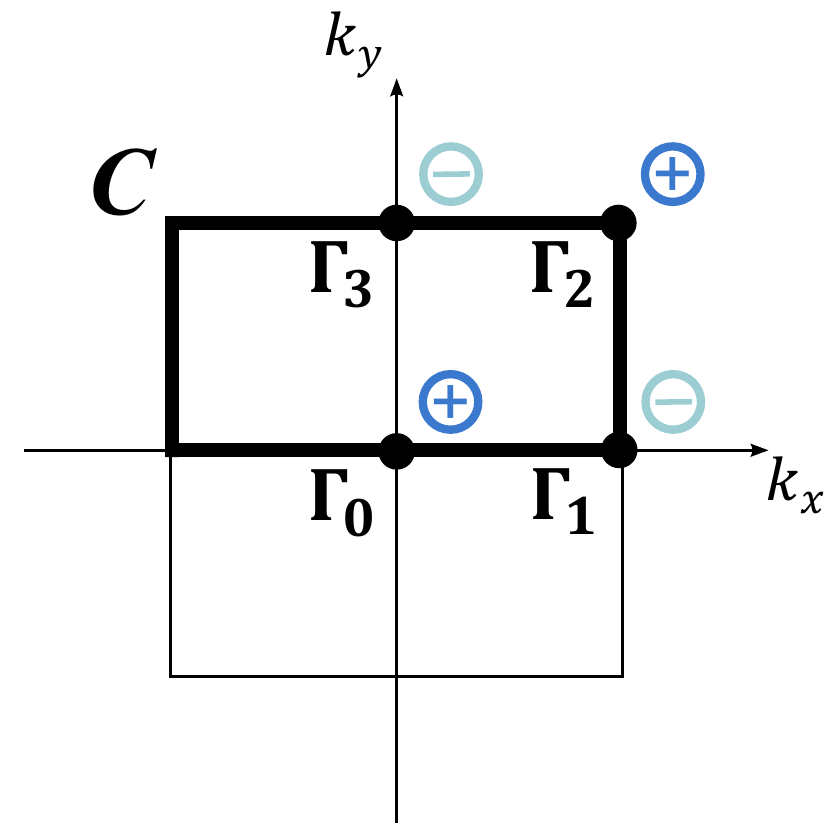}&
\includegraphics[trim = 18mm 12.5mm 5mm 0mm, clip, width=5.5cm]{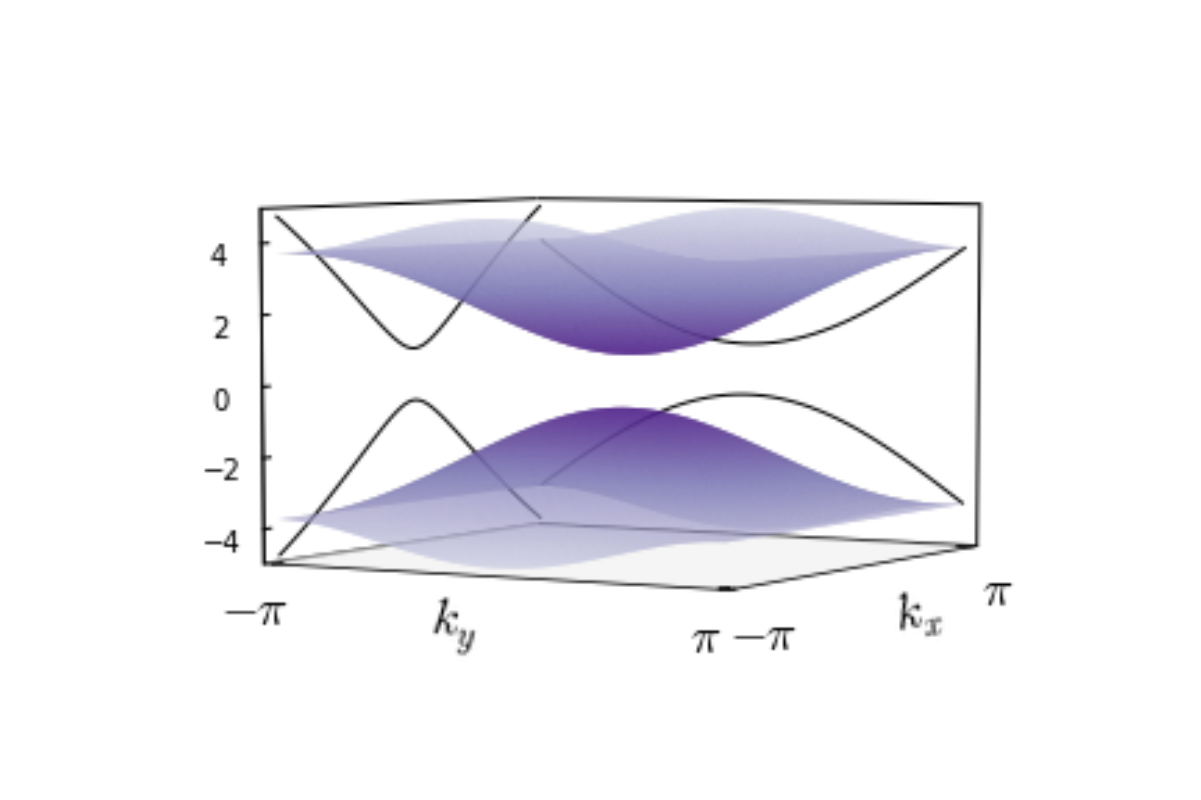}
\end{array}$
\caption{\small (Color online) Parity products at the time-reversal invariant momenta of the Brillouin zone (left) and lowest energy bands of phosphorene (right). Energy is given in $eV$.}
\label{Brillouin Zone}
\end{figure}
The first order in $\lambda$ refers to the time-averaged Hamiltonian, since electrons cannot follow the field dynamics. It implies that $\tilde H_{1}({\bf k})$ only differs from matrix (\ref{Bloch Hamiltonian Matrix}) in the field-renormalized hopping amplitudes. The second order in $\lambda$ describes the sum of all virtual emissions and absorptions of $m$ photons. A nearest-neighbor approximation yields $\tilde H_{2}({\bf k})=0$ and $\tilde H_{2}({\bf k}) = \mathcal{M}({\bf k}) \otimes \sigma_{3}$ for linear and elliptic polarizations, respectively \cite{SM}.
Therefore, elliptically polarized fields yield a non-vanishing second-order contribution that breaks time-reversal symmetry, as $\mathcal{M}^{*}(-{\bf k})=-\mathcal{M}({\bf k})$.\\

{\it{Topological transitions from band parities}} --- 
The purpose of the upcoming lines is to apprehend the topology of the photon-renormalized band structure of phosphorene given by $\tilde H({\bf k}) \simeq \lambda \tilde H_{1}({\bf k}) + \lambda^{2}  \tilde H_{2}({\bf k})$. To do so, we give a prescription in the spirit of Ref. \cite{Fu:2007hl} which relates the inversion eigenvalues at the time-reversal invariant points to the Berry phase of the occupied Bloch wavefunctions along half the Brillouin zone. Note that, if the occupation distribution is \textit{a priori} involved in the observation of nonequilibrium topological properties, transport is well described by the effective static band structure in the high-frequency limit, because electrons cannot absorb photons of the off-resonant light \cite{kitagawa2011transport}.

First, phosphorene is invariant under spatial inversion. This transformation consists of reversing all real-space coordinates, and interchanging electronic orbitals $A_{1}$ with $B_{1}$ and $A_{2}$ with $B_{2}$. In momentum space, this implies ${\cal{I}}^{\dagger} \tilde{H}({\bf k}) {\cal{I}} = \tilde{H}(-{\bf k})$, where the inversion operator is written in terms of Pauli matrices as ${\cal{I}} = \sigma_{1} \otimes \sigma _{1}$. At the time-reversal invariant momenta ${\bf \Gamma}={\bf G}/2$, this yields the commutation relation $[\tilde{H}({\bf \Gamma}),~ {\cal{I}}]=0$. Consequently, there exists a basis of Bloch eigenstates that are also eigenvectors of ${\cal I}$ with eigenvalues $\pi({\bf \Gamma})=\pm1$. This eigenvalue, named parity, labels the energy bands at time-reversal invariant momenta ${\bf \Gamma}_{i=0,...3}$ of the Brillouin zone defined in Fig.~\ref{Brillouin Zone}. Since we are interested in crossing between valance and conduction bands, we introduce the parity product of occupied energy bands indexed by $n$: $\delta_{i} = \prod_{n} \pi_{n}({\bf \Gamma}_{i})$. This quantity only changes when the crossing involves valence and conduction bands of opposite parities, which defines a band inversion. Besides, no band inversion can occur at ${\bf \Gamma}_{i=1,3}$ and $\delta_{i=1,3} = -1$ is fixed \cite{SM}. At momenta ${\bf \Gamma}_{i=0,2}$, this product is
\begin{align}\label{Parity Product}
\delta_{i} &= \sgn \left[ \Delta_{-}\left({\bf \Gamma_{i}}\right)\right] \sgn \left[\Delta_{+}\left({\bf \Gamma_{i}}\right) \right]  ~,
\end{align}
where $\Delta_{\pm}\left({\bf \Gamma_{i}}\right) =F_{2}\left({\bf \Gamma_{i}}\right) \pm F_{1}\left({\bf \Gamma_{i}}\right)$ refer to the energy gaps of the two occupied bands at ${\bf \Gamma_{i}}$.

\begin{figure}[t!]
\centering
$\begin{array}{cc}
\includegraphics[trim = 25mm 0mm 30mm 0mm, clip, width=4cm]{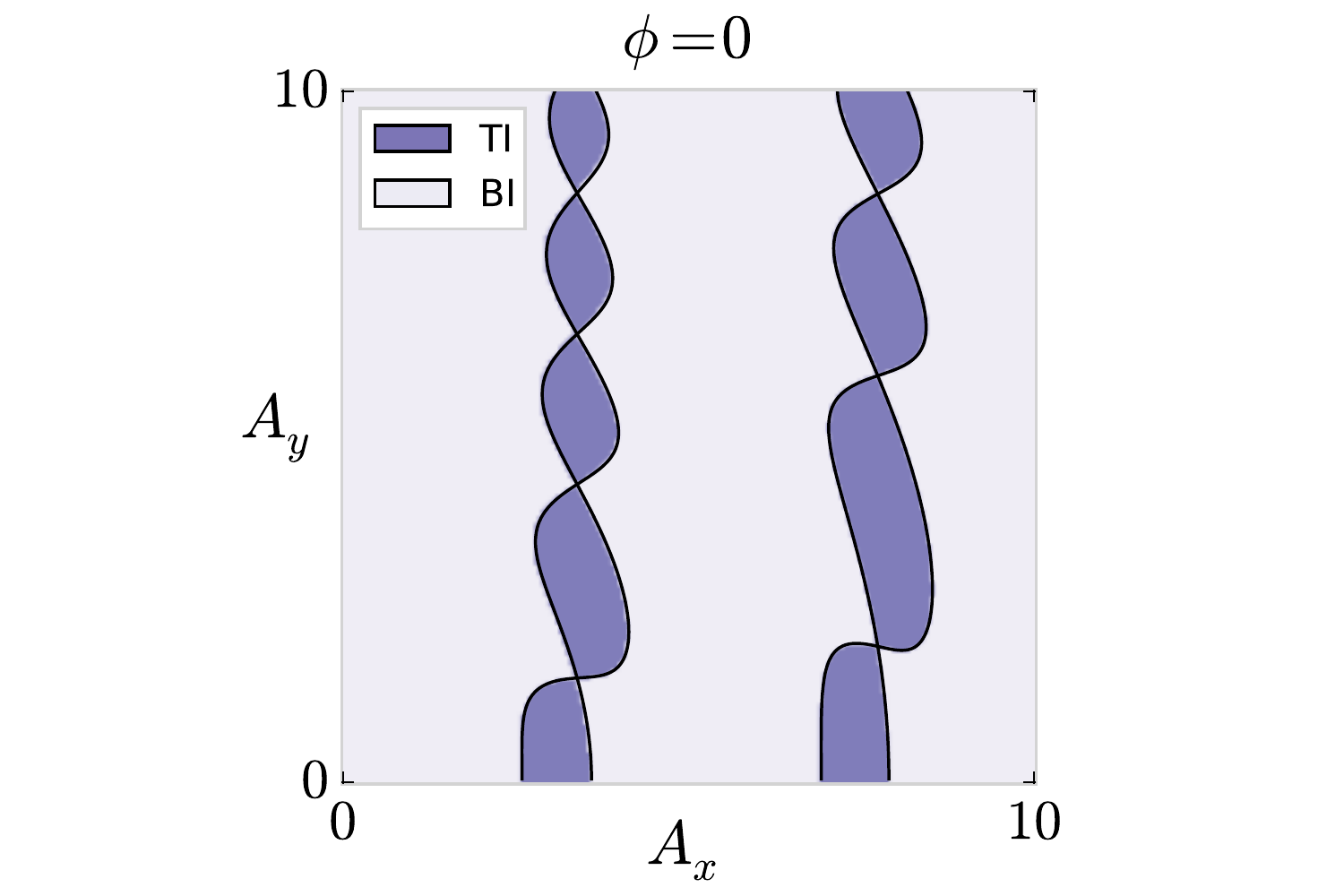}&
\includegraphics[trim = 25mm 0mm 30mm 0mm, clip, width=4cm]{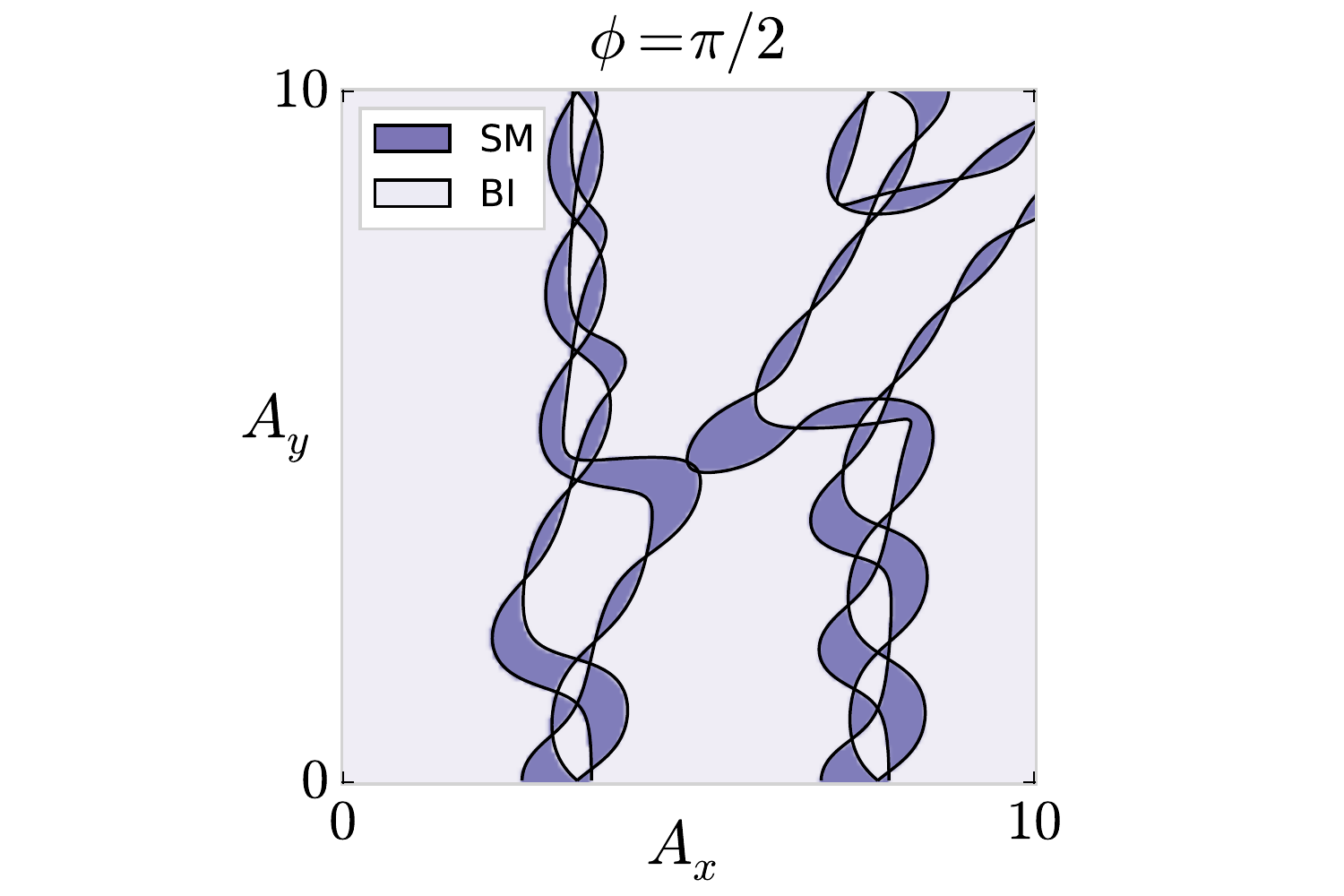}\\
\end{array}$
\caption{\small (Color online) Phase diagrams for electric fields with elliptic (left) and linear (right) polarizations. Light purple areas refer to band insulating (BI) phases characterized by $e^{i\gamma_{\mathcal{C}}}=+1$. Dark purple areas correspond to semimetallic (SM) and topological insulating (TI) phases in which $e^{i\gamma_{\mathcal{C}}}=-1$. Components $A_{x}$ and $A_{y}$ are given in $\AA^{-1}$.}
\label{Phase Diagrams}
\end{figure}

On the other hand, elliptic polarized fields break both time-reversal and chiral symmetries individually, meaning respectively that $\tilde H_{2}({\bf k})\neq\tilde H_{2}^{*}(-{\bf k})$ and $\mathcal{S} \tilde H_{2}({\bf k}) \mathcal{S} \neq -\tilde H_{2}({\bf k})$ with $\mathcal{S}=\sigma_{0}\otimes\sigma_{3}$. Nonetheless, their product always leads to the particle-hole symmetry $\mathcal{S}^{\dagger} \tilde H({\bf k}) \mathcal{S} = -\tilde H^{*}(-{\bf k})$, regardless of the field polarization. Thus, eigenstates come in pairs with opposite energies, and opposite parities because $\{ \mathcal{I}, \mathcal{S} \}=0$. The spectrum is then particle-hole symmetric and band inversions can only occur at zero energy. 
Based on inversion and particle-hole symmetries, we can finally connect the Berry phase $\gamma_{\mathcal{C}}$ picked up by the occupied Bloch wavefunctions along a path $\mathcal{C}$ which encloses half the Brillouin zone, to the parity products at ${\bf \Gamma_{i=0,2}}$ \cite{SM}. This results in
\begin{align}\label{Berry Phase and Parity Products}
e^{i\gamma_{\mathcal{C}}} &= \delta_{0}\delta_{2} ~.
\end{align}
This quantity, which is illustrated in Fig. \ref{Phase Diagrams}, changes signs only when photon-induced band inversions occur at ${\bf \Gamma_{i=0,2}}$. \\

\begin{figure}[b]
\centering
$\begin{array}{cc}
\includegraphics[trim = 20mm 10mm 17mm 20mm, clip, width=4.0cm]{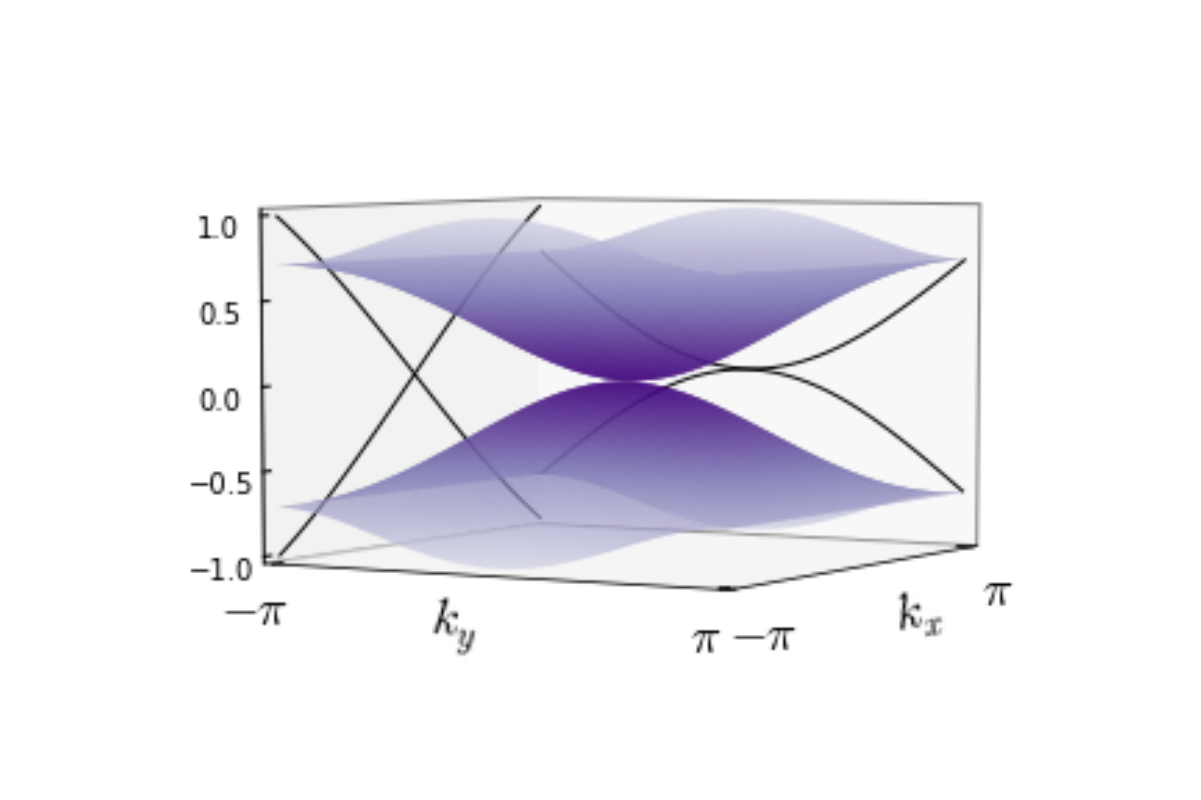}&
\includegraphics[trim = 20mm 10mm 17mm 20mm, clip, width=4.0cm]{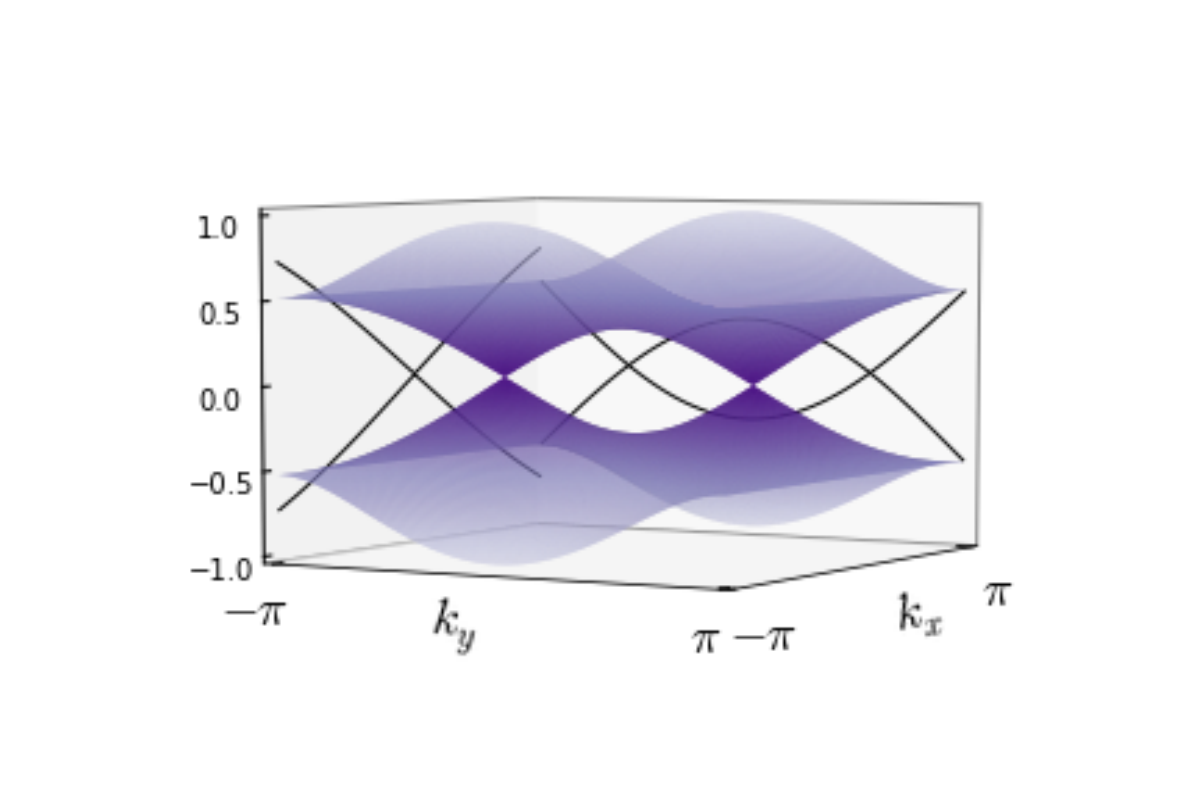}\\
\includegraphics[trim = 0mm 0mm 0mm 0mm, clip, width=3.2cm]{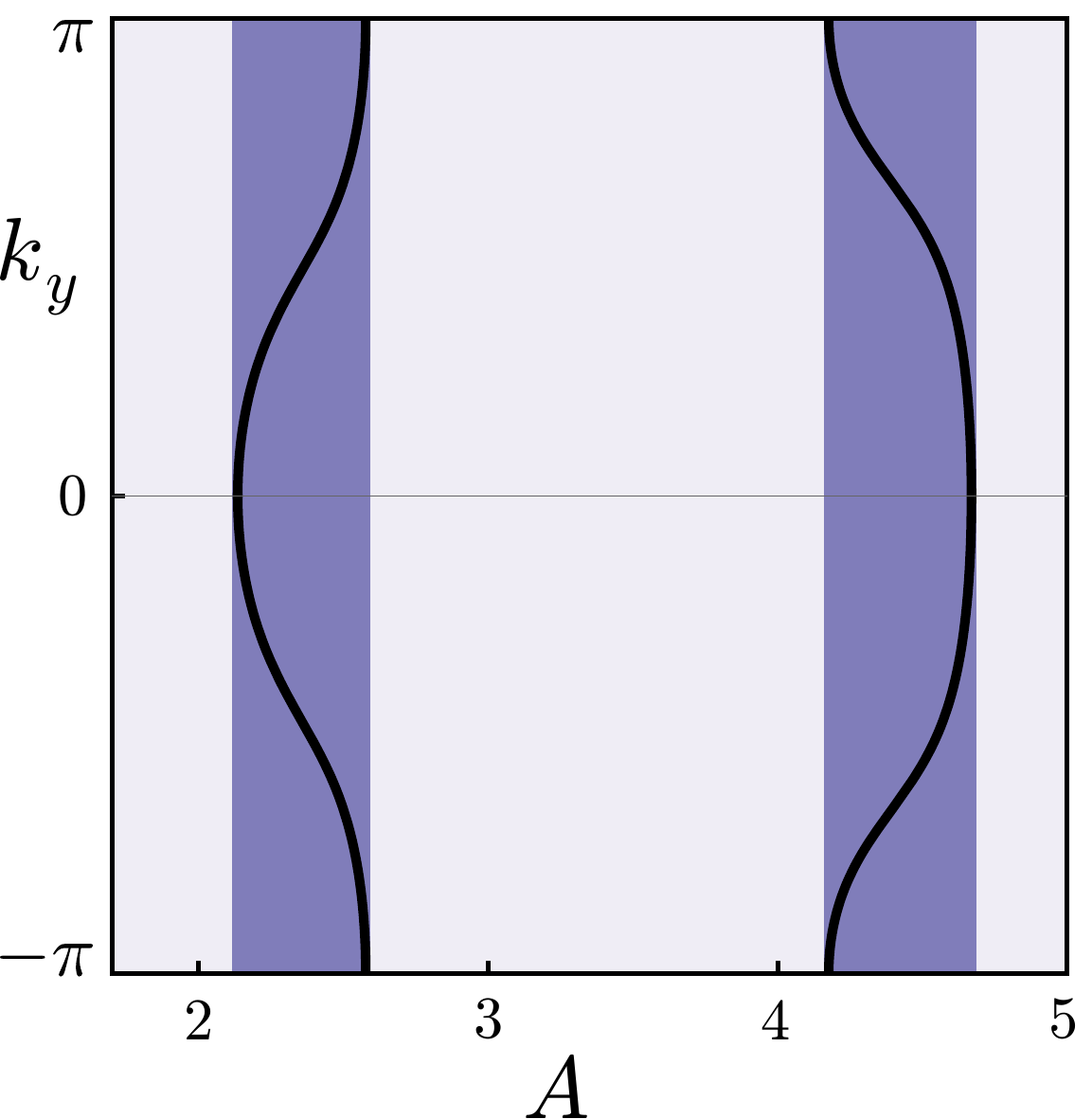}&
\includegraphics[trim = 0mm 0mm 0mm 0mm, clip, width=3.2cm]{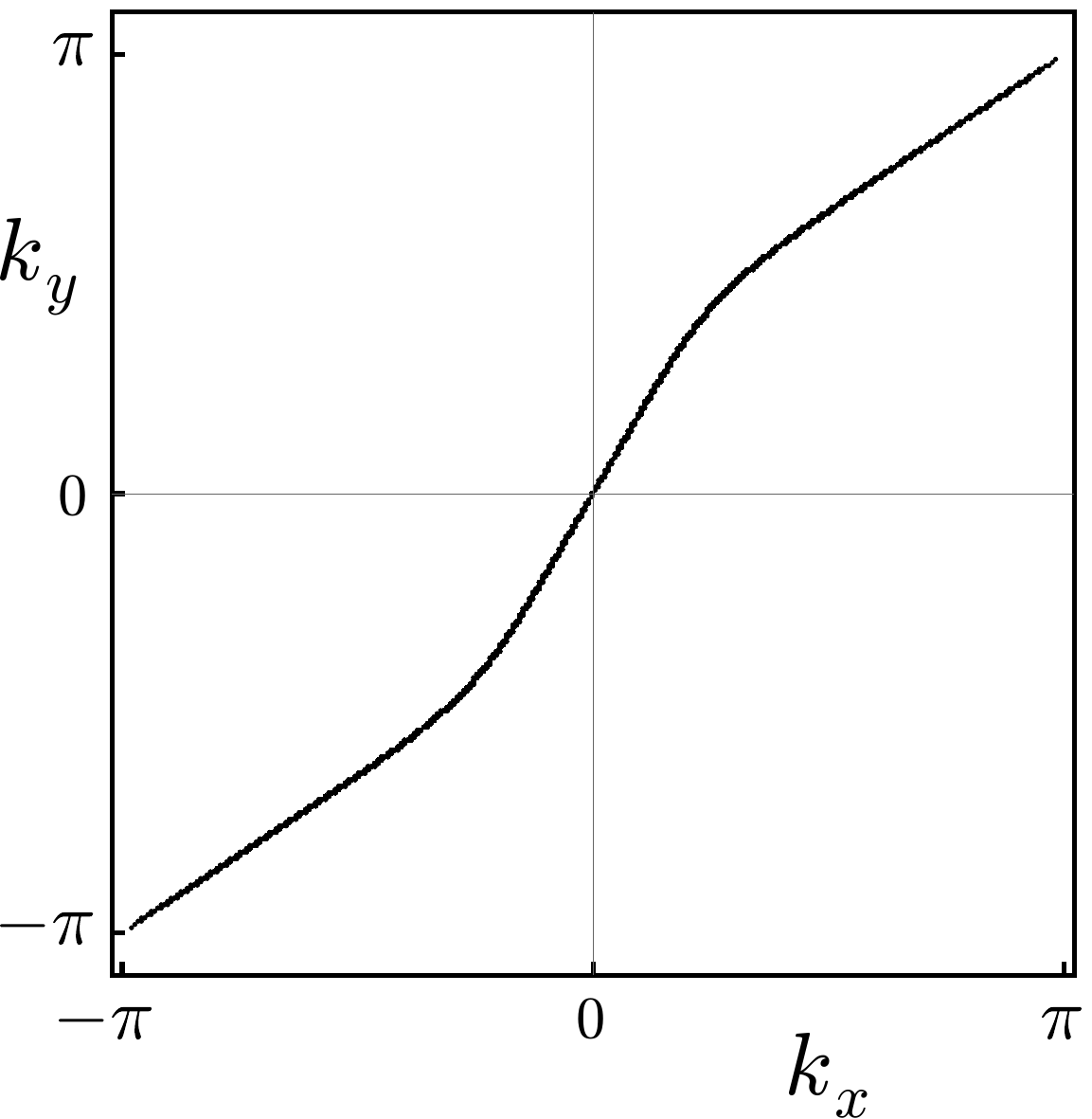}\\
\end{array}$
\caption{\small (Color online) Lowest-energy bands in the semimetallic phase induced by linearly polarized fields ($\phi=\pi/2$). Figures on top describe semi-Dirac (left) and massless Dirac electrons (right). Energy is given in eV. Figures at the bottom show the Dirac cones trajectories (black lines) for the configuration $A_{x}=A_{y}=A$.}
\label{Semimatallic Transition}
\end{figure}

{\it{Linear polarized fields}} --- 
For linear polarizations, phosphorene band-structure is simply described by the time average $\tilde H_{1}({\bf k})$ up to the second order in the high-frequency limit. Around ${\bf \Gamma_{0}}$, occupied energy bands ``$\pm$'' can be expanded as
\begin{align}
E_{\pm}({\bf \Gamma_{0}}+{\bf K}) &\simeq - \left|\Delta_{\pm}({\bf \Gamma_{0}})\right| - \frac{1}{2}~ {\bf K^{\text{T}}}\cdot \mathit{Hess}_{\pm}({\bf \Gamma_{0}})\cdot{\bf K} ~.
\end{align}
Here $K_{x}=~( k_{x}+~k_{y})/2$, $K_{y}=~(k_{x}-~k_{y})/2$, and Hessian matrix $\mathit{Hess}_{\pm}$ describes the local curvature of energy bands.  Energy gap $\Delta_{\pm}({\bf \Gamma_{0}})=~u_{1+}+~u_{1-}\pm~ u_{2}$ relies on field-renormalized hopping amplitudes $u_{1\pm}=~t_1\,J_0\left(\!\!\sqrt{\alpha_{x}^2+\alpha_{y}^2\pm2\alpha_{x}\alpha_{y}\sin\phi}\,\right)$ and $u_2= t_2\,J_0(\beta_{x})$, where $\alpha_{x}=|d_{1x}|\,A_{x}$, $\alpha_{y}=|d_{1y}|\,A_{y}$, and $\beta_{x}=|d_{3x}|\,A_{x}$. Besides, the nature of the dispersion relation is encoded in the determinant of the Hessian matrix, namely
\begin{align}
\Det \mathit{Hess}_{\pm}({\bf \Gamma_{0}}) = \pm \, u_{1+}u_{1-}u_{2} \sgn \left[ \Delta_{\pm}({\bf \Gamma_{0}})\right] ~.
\end{align}
In the absence of time-periodic field, hopping amplitudes fulfill $u_{1+}u_{1-}u_{2}>0$ and $\Delta_{\pm}({\bf \Gamma_{0}})>0$. Thus, $\Det \mathit{Hess}_{\pm}({\bf \Gamma_{0}})>0$ and the system lies in a band insulating phase, with elliptic parabolic dispersion relations. When tuning the field magnitude, $\Delta_{+}({\bf \Gamma_{0}})$ changes signs, and so does $\Det \mathit{Hess}_{+}({\bf \Gamma_{0}})$. It describes a hyperbolic parabolic dispersion relation, so that there are now two Dirac cones in the vicinity of ${\bf \Gamma_{0}}$. Therefore, the system has undergone a Lifshitz transition toward a semimetallic phase. From Eqs. (\ref{Parity Product}) and (\ref{Berry Phase and Parity Products}), this is equivalently characterized by $e^{i\gamma_{\mathcal{C}}}=-1$, where the non-vanishing Berry phase provides a topological stability of every Dirac cone inside $\mathcal{C}$. Right at the transition, when $\Delta_{+}=0$, the determinant of the Hessian matrix is null, whereas its trace is not: $\Tr \mathit{Hess}_{+}({\bf \Gamma_{0}})=u_{1+}^{2}+u_{1-}^{2}>0$. Consequently, the dispersion relation is linear in one direction albeit parabolic in others. This spectrum provides a universal signature of Dirac cone merging in two dimensions and is said to be semi-Dirac \cite{montambaux2009universal}. It happens for example for $A\sim2.1\,\AA^{-1}$ (see Fig.~\ref{Semimatallic Transition}), which can be obtained for magnitudes of the field larger than $30\,V/\AA$ . Here, one can show that, right at the transition, the semi-Dirac occupied band satisfies
\begin{align}
E_{-} ({\bf \Gamma_{0}}+{\bf Q}) &\simeq -\sqrt{ Q^{2}_{x}c^{2}_{x} + Q^{4}_{y}/4m^{2}} ~,
\end{align}
which refers to Dirac electrons with celerity $c_{x}=\sqrt{u_{1+}^{2}+u_{-1}^{2}}$ along $Q_{x}=\frac{u_{1+}K_{x}+u_{1-}K_{y}}{\sqrt{u_{1+}^{2}+u_{1-}^{2}}}$, and Shr\"odinger electrons with effective mass $m=\frac{u_{1+}^{2}+u_{-1}^{2}}{(u_{1+}+u_{-1})u_{+1}u_{-1}}$ along $Q_{y}=\frac{u_{1+}K_{y}-u_{1-}K_{x}}{\sqrt{u_{1+}^{2}+u_{1-}^{2}}}$.

A similar Lifshitz transition can also take place at ${\bf \Gamma_{2}}$, where the results above hold via the substitution $u_{1+}\to -u_{1+}$. Thus, the Dirac cones in semimetallic phase move along a line which connects momenta ${\bf \Gamma_{0}}$ and ${\bf \Gamma_{2}}$, where they can be created and annihilated in pairs. This is illustrated in Fig. \ref{Semimatallic Transition}. If some hopping amplitudes vanish, it is straightforward to see from Fig. \ref{Lattice} that electrons have a zero- or one-dimensional behavior which is unstable because it can be removed by perturbations such as distant hopping processes.\\

\begin{figure}[t]
\centering
$\begin{array}{cc}
\includegraphics[trim = 10mm 0mm 5mm 0mm, clip, width=4.2cm]{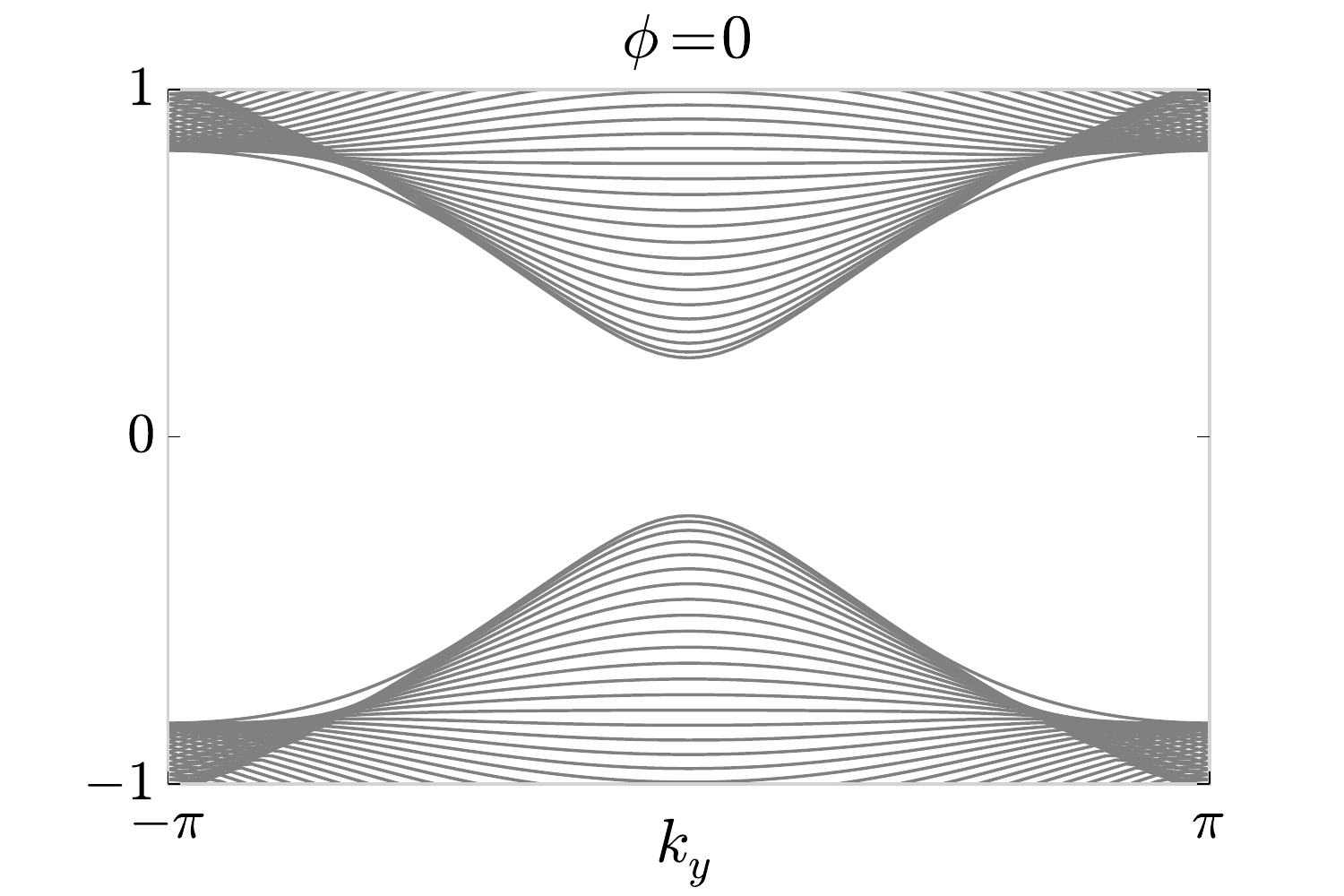}&
\includegraphics[trim = 10mm 0mm 5mm 0mm, clip, width=4.2cm]{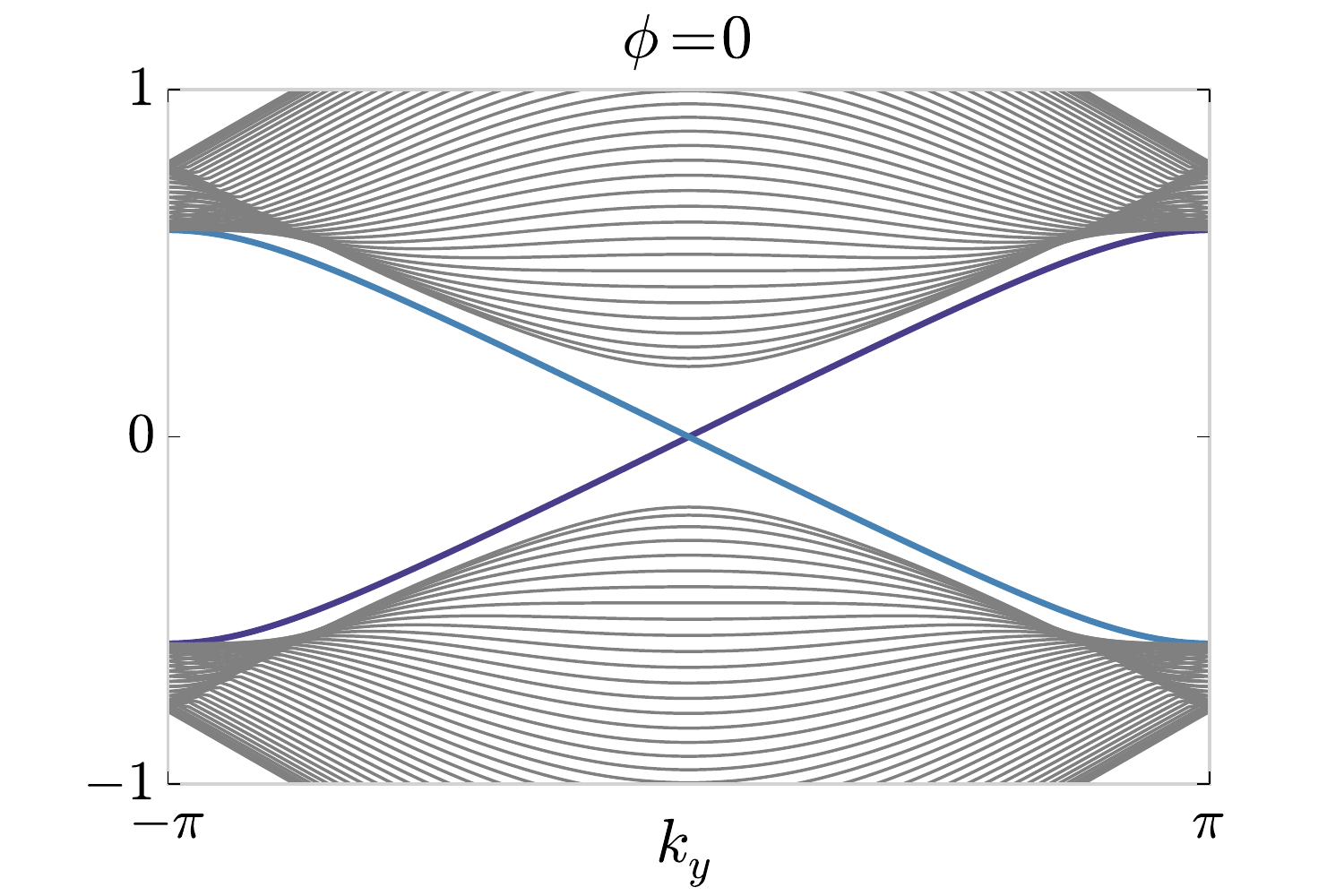}\\
\end{array}$
\caption{\small (Color online) Spectra of a ribbon with bearded termination along direction $y$, under circularly polarized fields $A_{x}=2.80\AA^{-1}$ (left) and $A_{x}=2.95\AA^{-1}$ (right). The latter yields chiral edge states within the bulk energy gap, in agreement with the topological phase predicted in Fig. \ref{Phase Diagrams}.}
\label{Bearded ribbon specta}
\end{figure}

{\it{Elliptic polarized fields}} --- 
For elliptic polarizations, the second order term described by $\lambda^{2}\mathcal{M}({\bf k}) \otimes \sigma_{3}$ is responsible for an energy gap which can only close at ${\bf \Gamma_{0}}$ \cite{SM}. Moreover, the Bloch band structure still has particle-hole symmetry, with respect to operator $\mathcal{S}$. It satisfies $\mathcal{S}^{2}=+1$, so that the system belongs to the Bogoliubov-de Gennes class D and its band-structure topology is characterized by a first Chern number $\nu$ in 2D \cite{PhysRevB.78.195125}. When neither of the energy gaps $\Delta_{\pm}({\bf \Gamma_{0}})$ vanish, we provide a connection in \cite{SM} between this topological invariant and the Berry phase in Eq. (\ref{Berry Phase and Parity Products}). This leads to $(-1)^{\nu}=\delta_{0}$.
When $\delta_{0}=-1$, the Chern number is odd and the system necessarily lies in a topological insulating phase. It guaranties the existence of particle-hole-symmetry-protected chiral boundary modes within the bulk energy gap, as illustrated in the case of a phosphorene ribbon with bearded termination in Fig. \ref{Bearded ribbon specta}. Moreover, it can easily be checked from \cite{SM} that the second order term $\lambda^{2}\mathcal{M}({\bf k}) \otimes \sigma_{3}$ effectively describes anisotropic next nearest-neighbor processes. Thus, $\tilde H({\bf k}) \simeq \lambda \tilde H_{1}({\bf k}) + \lambda^{2}  \tilde H_{2}({\bf k})$ is a dynamical realization of an anisotropic four-band Haldane's model with vanishing onsite potential and $\pi/2$ flux \cite{PhysRevLett.61.2015}, similarly to what was pointed out for graphene under circular polarized light \cite{kitagawa2011transport}. Consequently, the system has been driven into an anomalous quantum Hall phase.\\

In summary, we have described the stroboscopic dynamics of rapidly driven phosphorene up to the second order in the high-frequency limit. We gave a general prescription to apprehend the particle-hole-symmetry-protected topology of the photon-renormalized band-structure from parities of occupied Bloch wavefunctions. It enabled us to show that phosphorene is an anisotropic semiconductor in which engineering topologically stable chiral Dirac charge carriers, either two- or one-dimensional in the bulk, or one-dimensional along a boundary, may be achieved by tuning the polarization and intensity of a single external field. All these features occurring in the same non-compound thin film make it rather appealing for electronic nanodevice realizations.
\\

\begin{acknowledgments}
The authors would like to thank A. N. Rudenko, A. Itin, J. Mentink and S.~ Brener for useful comments. This work was supported by NWO via Spinoza Prize and by ERC Advanced Grant 338957 FEMTO/NANO.
\end{acknowledgments}

\bibliographystyle{apsrev4-1}
\bibliography{references}

\newpage
~
\newpage
\onecolumngrid

\section{Supplemental Material: Laser--induced topological transitions in phosphorene with inversion symmetry}

\subsection{First order in the high-frequency limit}

\subsubsection{Effective tight-binding description}

The first order in the high-frequency limit shows that the system can be described by the time-independent Hamiltonian $\tilde H_{1}=H_{0}$, where $H_{0}=\int_{-\pi}^{+\pi} \frac{dt}{2\pi} H(t)$ is the time-average of the Hamiltonian over a period. In momentum space, its derivation involves the calculation of terms like 
\begin{align*}
&\int_{-\pi}^{+\pi}\frac{dt}{2\pi}e^{i\,{\bf d_{x}}A_{x}\cos{}t}e^{i\,{\bf d_{y}}A_{y}\sin(t-\phi)}=J_0\left(\!\sqrt{(d_{x}A_{x})^2+(d_{y}A_{y})^2+2d_{x}d_{y}A_{x}A_{y}\sin\phi}\,\right) ~,
\end{align*}
which essentially results in the renormalization of hopping amplitudes. It is then straightforward to show that
\begin{align}
\tilde H_{1}({\bf k}) = 
\left( \begin{array}{llll} 
F_{3}({\bf k}) & F_{4}({\bf k}) & F_{1}({\bf k}) & F_{2}({\bf k}) \\
F_{4}^{*}({\bf k}) & F_{3}({\bf k}) & F_{2}({\bf k})~ e^{i\varphi({\bf k})} & F_{1}({\bf k}) \\
F_{1}^{*}({\bf k}) & F_{2}^{*}({\bf k})~ e^{-i\varphi({\bf k})} & F_{3}({\bf k}) & F_{4}({\bf k}) \\
F_{2}^{*}({\bf k}) & F_{1}^{*}({\bf k}) & F_{4}^{*}({\bf k}) & F_{3}({\bf k})
\end{array} \right) \notag ~,
\end{align}
where $F_1({\bf k}) = f_1+f_5+f_8$, $F_2({\bf k}) = f_2+f_6+f_9$, $F_3({\bf k}) = f_3+f_7+f_{10}$ and $F_4({\bf k}) = f_4$, and where functions $f_{i=1...10}$ are given by
\begin{flalign*}
f_1 &= t_1\,J_0\left(\sqrt{\alpha_{x}^2+\alpha_{y}^2-2\alpha_{x}\alpha_{y}\sin\phi}\right)+t_1\,J_0\left(\sqrt{\alpha_{x}^2+\alpha_{y}^2+2\alpha_{x}\alpha_{y}\sin\phi}\right)e^{ik_{y}}~, &\nonumber
\end{flalign*}
\begin{flalign*}
f_2 &= t_2\,J_0(\beta_{x})\,e^{-i\varphi({\bf k})}~, & \nonumber
\end{flalign*}
\begin{flalign*}
f_3 &= 2\,t_3\,J_0(2\alpha_{y})\cos{}k_{y}~, & \nonumber
\end{flalign*}
\begin{flalign*}
f_4 &= 2\,t_4\,J_0\left(\sqrt{(\alpha_{x}+\beta_{x})^2+\alpha_{y}^2+2(\alpha_{x}+\beta_{x})\alpha_{y}\sin\phi}\right)e^{-i\frac{\varphi({\bf k})}{2}}\cos{}K_{x}+2\,t_4\,J_0\left(\sqrt{(\alpha_{x}+\beta_{x})^2+\alpha_{y}^2-2(\alpha_{x}+\beta_{x})\alpha_{y}\sin\phi}\right)e^{-i\frac{\varphi({\bf k})}{2}}\cos{}K_{y}~, & \nonumber
\end{flalign*}
\begin{flalign*}
f_5 &= t_5\,J_0\left(\sqrt{(\alpha_{x}+2\beta_{x})^2+\alpha_{y}^2+2(\alpha_{x}+2\beta_{x})\alpha_{y}\sin\phi}\right)e^{-ik_{x}} +t_5\,J_0\left(\sqrt{(\alpha_{x}+2\beta_{x})^2+\alpha_{y}^2-2(\alpha_{x}+2\beta_{x})\alpha_{y}\sin\phi}\right)e^{-i\varphi({\bf k})}~, & \notag
\end{flalign*}
\begin{flalign*}
f_6 &= t_6\,J_0(2\alpha_{x}+\beta_{x})e^{ik_{y}}~, & \nonumber
\end{flalign*}
\begin{flalign*}
f_7 &= 2\,t_7\,J_0(2\alpha_{x}+2\beta_{x})\cos{}k_{x}~, & \nonumber
\end{flalign*}
\begin{flalign*}
f_8 &= t_8\,J_0\left(\sqrt{\alpha_{x}^2+9\alpha_{y}^2+6\alpha_{x}\alpha_{y}\sin\phi}\right)e^{2ik_{y}}+t_8\,J_0\left(\sqrt{\alpha_{x}^2+9\alpha_{y}^2-6\alpha_{x}\alpha_{y}\sin\phi}\right)e^{-ik_{y}}~, & \nonumber
\end{flalign*}
\begin{flalign*}
f_9 &= t_9\,J_0\left(\sqrt{(2\alpha_{x}+\beta_{x})^2+4\alpha_{y}^2+4(2\alpha_{x}+\beta_{x})\alpha_{y}\sin\phi}\right)e^{2ik_{y}} +t_9\,J_0\left(\sqrt{(2\alpha_{x}+\beta_{x})^2+4\alpha_{y}^2-4(2\alpha_{x}+\beta_{x})\alpha_{y}\sin\phi}\right)~, & \nonumber
\end{flalign*}
\begin{flalign*}
f_{10} &= 2\,t_{10}\,J_0\left(\sqrt{4(\alpha_{x}+\beta_{x})^2+4\alpha_{y}^2+8(\alpha_{x}+\beta_{x})\alpha_{y}\sin\phi}\right)\cos2K_{x} +2\,t_{10}\,J_0\left(\sqrt{4(\alpha_{x}+\beta_{x})^2+4\alpha_{y}^2-8(\alpha_{x}+\beta_{x})\alpha_{y}\sin\phi}\right)\cos2K_{y}~, &
\end{flalign*}
with $\alpha_{x}=|d_{1\,x}|\,A_{x}$, $\alpha_{y}=|d_{1\,y}|\,A_{y}$, $\beta_{x}=|d_{3\,x}|\,A_{x}$ and $K_{x}=~(k_{x}+~k_{y})/2$, $K_{y}=(k_{x}-k_{y})/2$. The expressions above rely on the tight-binding model introduced in \cite{Rudenko:2015kq}, where hopping parameters, given in $eV$, have been estimated as $t_{1}=-1.486$, $t_{2}=3.729$, $t_{3}=-0.252$, $t_{4}-0.071=$, $t_{5}=-0.019$, $t_{6}=0.186$, $t_{7}=-0.063$, $t_{8}=0.101$, $t_{9}=-0.042$ and $t_{10}=0.073$. It is worth mentioning here that the nearest-neighbor hopping amplitudes are at least one order of magnitude larger than further hopping amplitudes.

\subsubsection{Parity products of occupied energy bands}

\begin{figure}[b]
\centering
$\begin{array}{cc}
\includegraphics[trim = 0mm 0mm 0mm 0mm, clip, width=7cm]{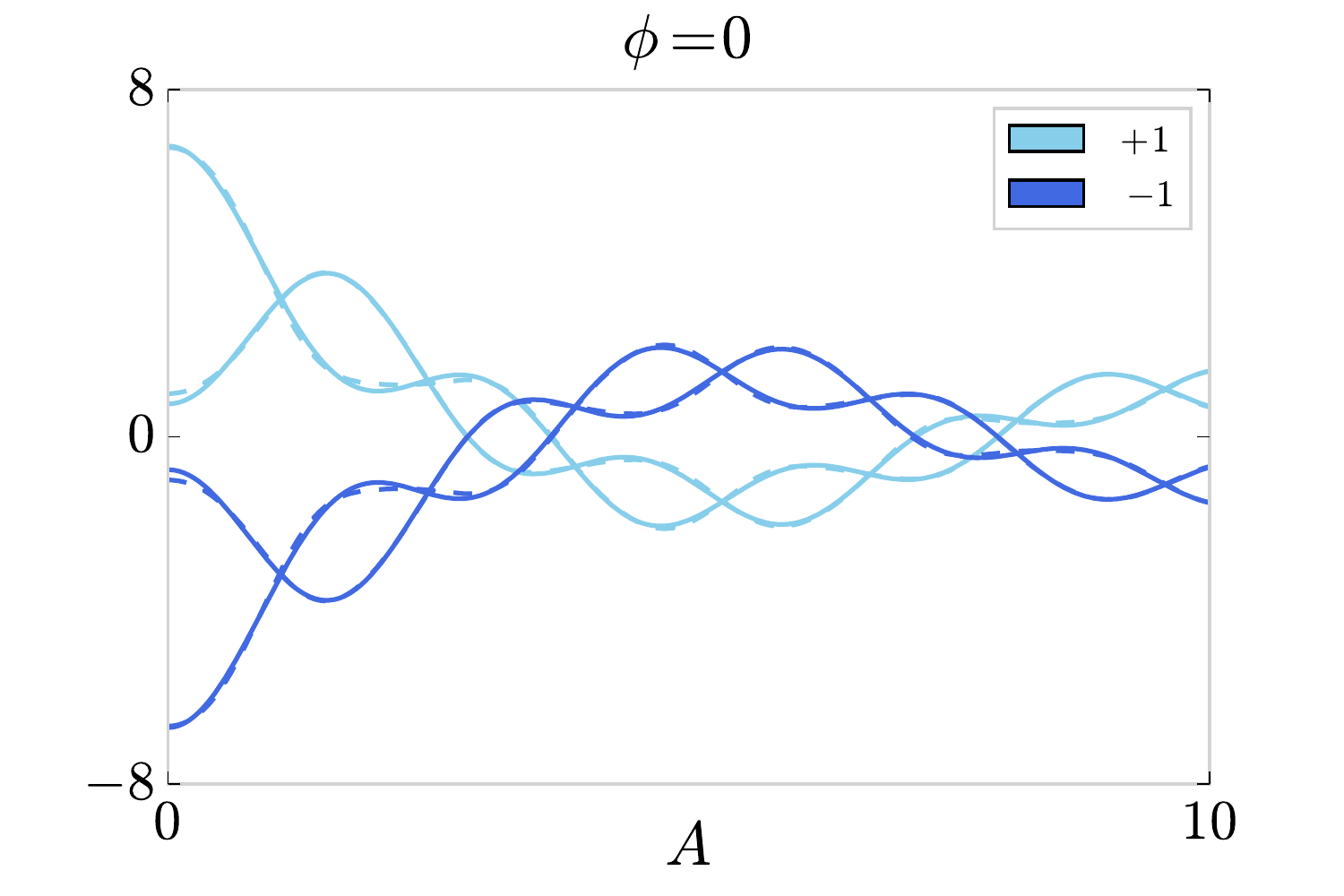}&
\includegraphics[trim = 0mm 0mm 0mm 0mm, clip, width=7cm]{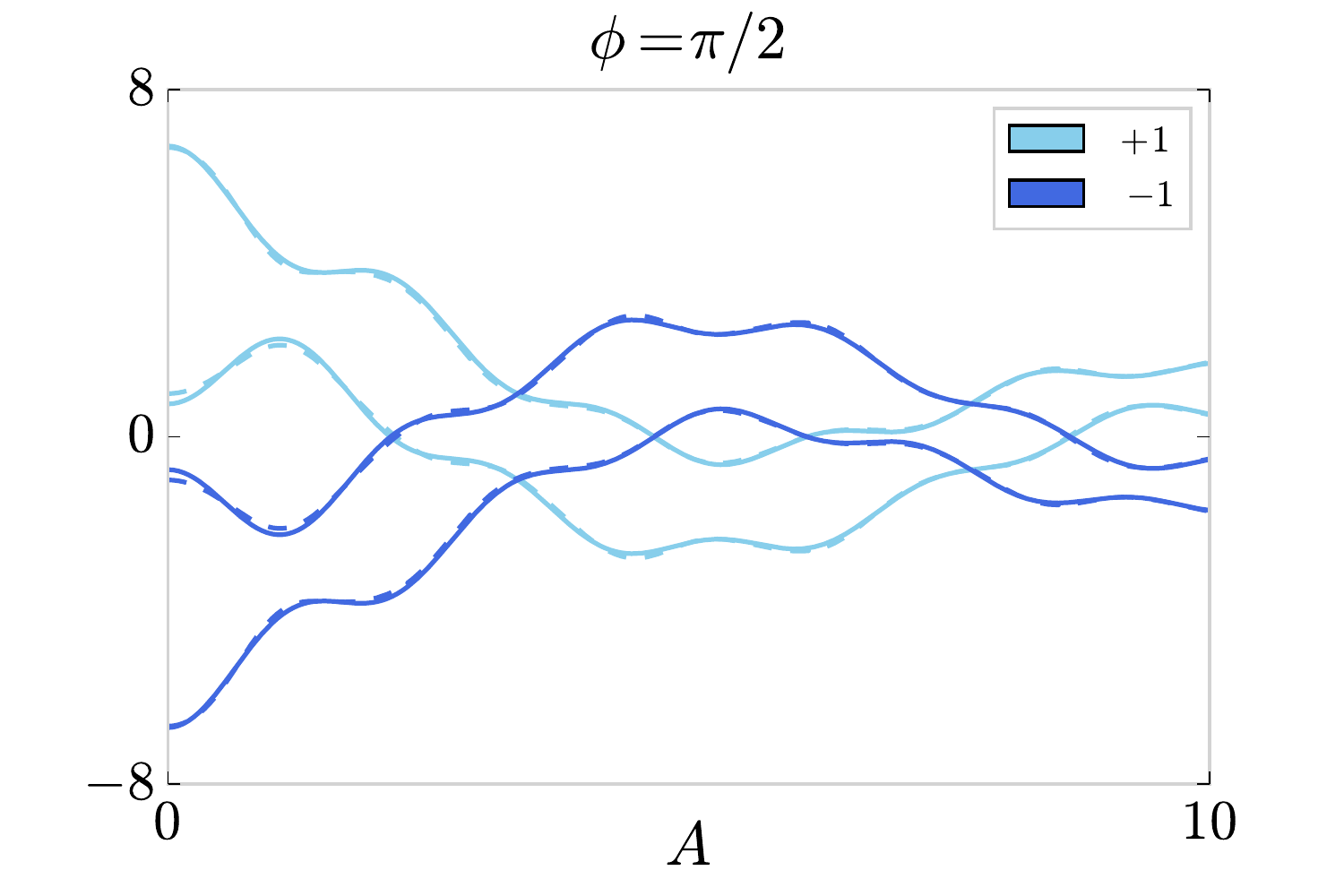}
\end{array}$
\caption{\small (Color online) Illustration of the four energy levels at ${\bf \Gamma_{0}}$ for elliptic ($\phi=0$) and linear ($\phi=\pi/2$) polarizations as a function of $A_{x}=A_{y}=A$, given in $\AA^{-1}$. In both cases, $F_{3}$ has been taken as the energy origin and energy is given in $eV$. The dashed lines refer to the ten-hopping model and are in good agreement with the full ones that refer to the nearest-neighbor approximation. Light and dark blues refer to positive ($+1$) and negative ($-1$) parities, respectively}
\label{AppendixBandInversionGamma0}
\end{figure}

As long as the phosphorene crystal has inversion symmetry, the Bloch Hamiltonian matrix commutes with the inversion operator $\mathcal{I}=\sigma_{1}\otimes\sigma_{1}$ at the time-reversal invariant momenta, i.e. $\left[ \mathcal{I}, \tilde H({\bf \Gamma_{i}}) \right]=0$. Thus, the eigenvalues of $\mathcal{I}$, namely $\pi_{n}=\pm1$, become good quantum numbers that label the energy bands. At momenta ${\bf \Gamma_{0}}$ and ${\bf \Gamma_{2}}$, the four energy bands satisfy
\begin{align}
\mathcal{E}_{1}({\bf \Gamma_{0,2}}) &= F_{3} + F_{2} - F_{1} - F_{4} ~~~\text{with}~~~ \pi_{1}({\bf \Gamma_{0,2}})=+1, \notag \\
\mathcal{E}_{2}({\bf \Gamma_{0,2}}) &= F_{3} + F_{2} + F_{1} + F_{4} ~~~\text{with}~~~ \pi_{2}({\bf \Gamma_{0,2}})=+1, \notag \\
\mathcal{E}_{3}({\bf \Gamma_{0,2}}) &= F_{3} - F_{2} - F_{1} + F_{4} ~~~\text{with}~~~ \pi_{3}({\bf \Gamma_{0,2}})=-1, \notag \\
\mathcal{E}_{4}({\bf \Gamma_{0,2}}) &= F_{3} - F_{2} + F_{1} - F_{4} ~~~\text{with}~~~ \pi_{4}({\bf \Gamma_{0,2}})=-1. \notag
\end{align}
Although we voluntarily omit to mention it in the expressions above for more clarity, the momentum dependent functions $F_{j=1,...4}$ are evaluated at time-reversal invariant points. Besides, the order of magnitudes of hopping amplitudes ($t_{1,2}\gg t_{i=3...10}$) implies that inversions between valence and conduction bands are likely to occur only between bands of opposite parities, either when $\mathcal{E}_{2}({\bf \Gamma_{0,2}})=\mathcal{E}_{3}({\bf \Gamma_{0,2}})$ or when $\mathcal{E}_{1}({\bf \Gamma_{0,2}})=\mathcal{E}_{4}({\bf \Gamma_{0,2}})$. Both conditions are independent of $F_{3}$ and $F_{4}$, and can be rewritten as
\begin{align}
F_{2}\pm F_{1} = 0. \notag
\end{align}
The parity product of the (two) occupied energy bands at ${\bf \Gamma_{i=0,2}}$ is then given by
\begin{align}
\delta_{i} &= \prod_{\text{occupied } n} \pi_n({\bf \Gamma_{i}}) \notag \\
&= \sgn\left[ F_{2}({\bf \Gamma_{i}}) - F_{1}({\bf \Gamma_{i}})\right] \sgn\left[ F_{2}({\bf \Gamma_{i}}) + F_{1}({\bf \Gamma_{i}})\right]. \notag
\end{align}
Although this criterion is valid within the ten hopping model, it suggests that band inversions are already well described within a nearest-neighbor tight-binding approximation as, first, distant processes are at least one order of magnitude smaller than nearest-neighbor ones and, second, their associated zeroth-order Bessel functions decay faster with the field than for nearest neighbors. This is illustrated in Fig. \ref{AppendixBandInversionGamma0} which depicts the four energy levels at ${\bf \Gamma_{0}}$ for elliptically and linearly polarized fields within both frameworks the ten hopping model and the nearest-neighbor approximation.

At momenta ${\bf \Gamma_{1}}$ and ${\bf \Gamma_{3}}$, one similarly obtains
\begin{align}
\mathcal{E}_{1}({\bf \Gamma_{1,3}}) &= F_{3} + \sqrt{F_{2}^{2} + \left(F_{1}+F_{4}\right)^{2}} ~~~\text{with}~~~ \pi_{1}({\bf \Gamma}_{1,3})=+1, \notag \\
\mathcal{E}_{2}({\bf \Gamma_{1,3}}) &= F_{3} + \sqrt{F_{2}^{2} + \left(F_{1}-F_{4}\right)^{2}} ~~~\text{with}~~~ \pi_{2}({\bf \Gamma}_{1,3})=-1, \notag \\
\mathcal{E}_{3}({\bf \Gamma_{1,3}}) &= F_{3} - \sqrt{F_{2}^{2} + \left(F_{1}-F_{4}\right)^{2}} ~~~\text{with}~~~ \pi_{3}({\bf \Gamma}_{1,3})=-1, \notag \\
\mathcal{E}_{4}({\bf \Gamma_{1,3}}) &= F_{3} - \sqrt{F_{2}^{2} + \left(F_{1}+F_{4}\right)^{2}} ~~~\text{with}~~~ \pi_{4}({\bf \Gamma}_{1,3})=+1. \notag
\end{align}
Importantly, when $\mathcal{E}_{2}({\bf \Gamma_{1,3}})=\mathcal{E}_{3}({\bf \Gamma_{1,3}})$, or when $\mathcal{E}_{1}({\bf \Gamma_{1,3}})=\mathcal{E}_{4}({\bf \Gamma_{1,3}})$, no band inversion can occur at ${\bf \Gamma_{1}}$ and ${\bf \Gamma_{3}}$, where the parity product of the occupied energy bands is fixed: $\delta_{1,3}=-1$. This results in
\begin{align}
\prod_{i=0}^{3}\delta_{i} &= \delta_{0}\delta_{2} ~. \notag
\end{align}
This quantity only depends on band inversions likely occuring at time reversal invariant momenta ${\bf \Gamma_{0}}$ or ${\bf \Gamma_{2}}$, a mechanism that plays a central role in topological features of phosphorene band-structure and will be the purpose of the upcoming sections. Note finally that, in the nearest-neighbor approximation, $F_{2}\pm F_{1}$ corresponds to the energy gap, hence the notation $\Delta_{\pm}({\bf \Gamma_{0,2}}) = F_{2}({\bf \Gamma_{0,2}}) - F_{1}({\bf \Gamma_{0,2}})$. The parity product of the occupied energy bands ${\bf \Gamma_{i=0,2}}$ can then be rewritten as
\begin{align}\label{Appendix Parity Product}
\delta_{i} &= \sgn\left[ \Delta_{+}({\bf \Gamma_{i}}) \right] \sgn\left[ \Delta_{-}({\bf \Gamma_{i}}) \right] ~.
\end{align}

\subsubsection{Band inversions from parity products}

In the high-frequency regime, the Bloch band structure is described by the time-average of the Hamiltonian matrix which, limited to nearest-neighbor processes, reads
\begin{align*}
\tilde H_{1}({\bf k}) = 
\left( \begin{array}{llll} 
0 & 0 & F_{1}({\bf k}) & F_{2}({\bf k}) \\
0 & 0 & F_{2}({\bf k})~ e^{i\varphi({\bf k})} & F_{1}({\bf k}) \\
F_{1}^{*}({\bf k}) & F_{2}^{*}({\bf k})~ e^{-i\varphi({\bf k})} & 0 & 0 \\
F_{2}^{*}({\bf k}) & F_{1}^{*}({\bf k}) & 0 & 0
\end{array} \right).
\end{align*}
Since it satisfies the relation ${\cal{S}} \tilde H_{1}({\bf k}) {\cal{S}} = -\tilde H_{1}({\bf k})$, 
where ${\cal{S}} =~ \sigma_{0} \otimes \sigma_{3}$, $\tilde H_{1}$ is said to have chiral symmetry.  As a result, the eigenstates come in pairs with opposite energies and the spectrum is particle-hole symmetric. One can then square the above matrix to show that the dispersion relations of the occupied energy bands labelled by the index $\pm$ satisfy
\begin{align}
\mathcal{E}_{\pm}^2 ({\bf K}) &= [u_{1+}\cos{}K_{x}+u_{1-}\cos{}K_{y}\pm{}u_2]^2 +[u_{1+}\sin{}K_{x}+u_{1-}\sin{}K_{y}\,]^2, \notag
\end{align}
where $u_{1\pm}=t_1\,J_0\left(\sqrt{\alpha_{x}^2+\alpha_{y}^2\pm2\alpha_{x}\alpha_{y}\sin\phi}\right)$ and $u_{2}=t_2\,J_0(\beta_{x})$. Band inversions between valence and conduction bands occur at zero energy and extrema of the dispersion relation. Those extrema fulfil $\nabla_{\bf k} \mathcal{E}^{2}_{\pm}=0$ and $\mathcal{E}_{\pm}\neq0$, or equivalently
\begin{align*}
\left \{
\begin{aligned}
0 &= u_{2}~ \big[ u_{1+}\sin{}K_{x} + u_{1-}\sin{}K_{y} \big] \\
0 &=\big[ 2~ u_{1-}\cos{}K_{y}\pm u_{2} \big]~u_{1+} \sin{}K_{x}- \big[ 2~ u_{1+}~\cos{}K_{x} 
\pm u_{2} \big]~u_{1-} \sin{}K_{y}.
\end{aligned}
\right .
\end{align*}
If $u_{1\pm}\neq0$ and $u_{2}\neq0$, then 
\begin{align*}
\left \{
\begin{aligned}
0 &= \big[ u_{1+}\sin{}K_{x} + u_{1-}\sin{}K_{y} \big] \\
0 &=\big[ u_{1+}~\cos{}K_{x}+ u_{1-}\cos{}K_{y}\pm u_{2} \big]~u_{1+} \sin{}K_{x},
\end{aligned}
\right .
\end{align*}
where $\mathcal{E}_{\pm} ({\bf K}) = [u_{1+}\cos{}K_{x}+u_{1-}\cos{}K_{y}\pm{}u_2]\neq0$. Thus, the extrema of the dispersion relation are necessarily located at ${\bf \Gamma_{0}}$ and ${\bf \Gamma_{2}}$. They are the only momenta where band inversions are allowed to take place. Consequently, the parity products $\delta_{0}$ and $\delta_{2}$ are sufficient to keep track of band inversions.

Note that other cases for which at least one of the parameters $u_{1\pm}$ or $u_{2}$ vanishes turn out to be unstable as, for example, they may be removed by distant hopping processes.

\subsection{Second order in the high-frequency limit}

\subsubsection{Effective tight-binding description}

As detailed in the main text, the second order in the high-frequency limit is described by the time-independent matrix:
\begin{align}
\tilde H_{2} &= -\sum_{m>0}\frac{[H_{m},H_{-m}]}{m} ~. \notag
\end{align}
Since the system is qualitatively well characterized within the nearest-neighbor tight-binding approximation, one finds that
\begin{align}
H_{m}({\bf k}) &= 
\left( \begin{array}{llll} 
0 & 0 & F_{1,m}({\bf k}) & F_{2,m}({\bf k}) \\
0 & 0 & F_{2,m}({\bf k})~ e^{i\varphi({\bf k})} & F_{1,m}({\bf k}) \\
F_{1,-m}^{*}({\bf k}) & F_{2,-m}^{*}({\bf k})~ e^{-i\varphi({\bf k})} & 0 & 0 \\
F_{2,-m}^{*}({\bf k}) & F_{1,-m}^{*}({\bf k}) & 0 & 0
\end{array} \right) ~, \notag
\end{align}
where
\begin{flalign}
F_{1,m}({\bf k}) &=
t_{1} \int_{-\pi}^{+\pi} \frac{dt}{2\pi} ~e^{imt} ~e^{ia_{x}\cos t} ~e^{ia_{y}\sin (t-\phi)}+ t_{1} ~e^{ik_{y}} \int_{-\pi}^{+\pi} \frac{dt}{2\pi} ~e^{imt} ~e^{ia_{x}\cos t} ~e^{-ia_{y}\sin (t-\phi)} & \notag \\
&= t_{1} ~e^{-im\pi/2}~\left[ e^{im\theta_{1}} ~J_{m}\left(\sqrt{a_{x}^{2}+a_{y}^{2}+2a_{x}a_{y}\cos(\phi+\pi/2)} \right) + e^{im\theta_{2}} ~J_{m}\left(\sqrt{a_{x}^{2}+a_{y}^{2}-2a_{x}a_{y}\cos(\phi+\pi/2)} \right) ~e^{ik_{y}}\right] ~, & \notag
\end{flalign}
and
\begin{flalign}
F_{2,m}({\bf k}) &= t_{2} ~e^{-im\pi/2} J_{m}(c) ~e^{-i\varphi({\bf k})} ~. & \notag
\end{flalign}
In the expressions above, it is implied that $\theta_{1} = - \arctan \frac{a_{y}\sin(\phi+\pi/2)}{a_{x}+a_{y}\cos(\phi+\pi/2)}$, $\theta_{2} = + \arctan \frac{a_{y}\sin(\phi+\pi/2)}{a_{x}-a_{y}\cos(\phi+\pi/2)}$, $a_{x}=d_{1x}A_{x}$ and $a_{y}=d_{1y}A_{y}$. Besides, one can show that
\begin{align}
H_{m}({\bf k})H_{-m}({\bf k}) = 
\left( \begin{array}{ll} 
 K_{m}({\bf k}) & 0 \\
 0 & K_{-m}({\bf k})
\end{array} \right) ~, ~~~\text{with}~~~
K_{m}({\bf k})= 
\left( \begin{array}{ll} 
F_{1,m}F^{*}_{1,m} + F_{2,-m}F^{*}_{2,-m} & F_{1,m}F^{*}_{2,m}e^{-i\varphi}+F_{2,-m}F^{*}_{1,-m} \\
F_{2,m}F^{*}_{1,m}e^{i\varphi}+F_{1,-m}F^{*}_{2,-m} & F_{1,m}F^{*}_{1,m} + F_{2,-m}F^{*}_{2,-m}
\end{array} \right) ~. \notag
\end{align}
The commutation relation can then be rewritten as $[H_{m},H_{-m}] = (K_{m}-K_{-m})\otimes \sigma_{3}$. Two cases have to be distinguished.
\begin{itemize}
\item For linear polarizations ($\phi=\pi/2$ and $\theta_{1}=\theta_{2}=0$),
\begin{align}
\tilde H_{2}& = 0 ~. \notag
\end{align}
As a result, the system is still described by its time-average. \textit{A fortiori} chiral and time-reversal symmetries, as well as invariance under spatial inversion are preserved.
\item For elliptic polarizations ($\phi=0$ and $\theta=\theta_{1}=\theta_{2}$),
\begin{align}
\tilde H_{2}({\bf k}) &= \tau_{3} \sin\left(k_{y}\right) ~ \sigma_{0}\otimes \sigma_{3} + \tau_{4} \cos\left(\frac{k_{x}}{2}\right) \sin\left(\frac{k_{y}}{2}\right)
\left[ \cos\left(\frac{k_{x}-k_{y}}{2}\right) \sigma_{1}\otimes \sigma_{3} +  \sin\left(\frac{k_{x}-k_{y}}{2}\right) \sigma_{2}\otimes \sigma_{3} \right] \notag ~,
\end{align}
where 
\begin{flalign}
&\left \{
\begin{aligned}
\tau_{3}&=4t^{2}_{1} \sum_{m>0} \frac{J^{2}_{m}\left(a\right)\sin(2m\theta)}{m} \\
\tau_{4}&=8t_{1}t_{2}\sum_{m>0} \frac{J_{m}\left(a\right)J_{m}\left(c\right)\sin(m\theta)}{m} ~.
\end{aligned}
\right . & \notag 
\end{flalign}
Though the system is still invariant under inversion symmetry, i.e. $\mathcal{I}^{\dagger}\tilde H_{2}({\bf k})\mathcal{I}=\tilde H_{2}(-{\bf k})$, time-reversal and chiral symmetries are both broken individually, since respectively $\tilde H^{*}_{2}(-{\bf k})=-\tilde H_{2}({\bf k})$ and $\mathcal{S}^{\dagger}\tilde H_{2}({\bf k})\mathcal{S}=\tilde H_{2}({\bf k})$.
\end{itemize}

\subsubsection{Band inversions from parity products}

The system is described by the time-independent effective Hamiltonian $\tilde H \simeq \lambda\tilde H_{1}+\lambda^{2} \tilde H_{2}$. In the case of a linearly polarized field, $\tilde H_{2}=0$. Consequently, the criterion already given in Eq. (\ref{Appendix Parity Product}) for the time average in order to keep track of band inversions still holds. In the case of elliptically polarized fields, however, $\tilde H_{2}\neq0$ and we have to find the momenta where band inversions are likely to occur. To do so, one can first remark that, although chiral and time-reversal symmetries are individually broken, their product leads to the following particle-hole symmetry relation:
\begin{align}
\mathcal{S}^{\dagger} \tilde H({\bf k}) \mathcal{S} = -\tilde H^{*}(-{\bf k}) ~. \notag
\end{align}
Thus, eigenstates come in pairs $|\psi_{n}\left({\bf k}\right)\rangle$ and $\mathcal{S}|\psi^{*}_{n}\left(-{\bf k}\right)\rangle$ with opposite energies, respectively $\mathcal{E}\left({\bf k}\right)$ and $-\mathcal{E}\left(-{\bf k}\right)$. The spectrum is particle-hole symmetric and band inversions can only occur at zero energy. In order to evaluate at what momenta they are likely to take place, we can simply solve $\Det \tilde H ({\bf k})=0$, which can equivalently be rewritten as
\begin{align*}
\left \{
\begin{aligned}
&u_{1}\cos\left(\frac{k_{y}}{2}\right) = \pm u_{2} e^{-i\frac{k_{x}}{2}}\\
&\tau_{3}\cos\left(\frac{k_{y}}{2}\right)\sin\left(\frac{k_{y}}{2}\right) = \pm \tau_{4} \cos\left(\frac{k_{x}}{2}\right)\sin\left(\frac{k_{y}}{2}\right) ~.
\end{aligned}
\right .
\end{align*}
When $u_{1}\tau_{4} \neq u_{2}\tau_{3}$, where $u_{1}=u_{1+}=u_{1-}$ in the case of elliptic polarization, the determinant only vanishes at ${\bf k}={\bf \Gamma_{0}}$ for $\Delta_{\pm}({\bf \Gamma_{0}}) =0$. As a result, elliptically polarized fields yield an energy gap which is only allowed to close at ${\bf \Gamma_{0}}$. This implies that $\delta_{2}=+1$ is fixed, and finally
\begin{align}
\prod_{i=0}^{3}\delta_{i} &= \delta_{0} ~. \notag
\end{align}

\subsection{Topological transitions from band inversions}

In this section, we first define some properties of the Berry connexion and the Berry curvature, before establishing their relation to topological features of the Bloch band structure, namely a $\pi$-quantized Berry phase and a first Chern number.

\subsubsection{Berry connection and Berry curvature}
Starting from its definition, the Berry connection associated to all bands (occupied and empty) satisfies
\begin{align}
\mathcal{A}\left({\bf k}\right) &= i \sum_{n} \langle \psi_{n}\left({\bf k}\right) | \nabla_{\bf k} | \psi_{n}\left({\bf k}\right) \rangle \notag \\
&= i \sum_{n} \langle \psi^{*}_{n}\left(-{\bf k}\right) |S^{\dagger} | \nabla_{\bf k}| S| \psi^{*}_{n}\left(-{\bf k}\right) \rangle \notag \\
&= i \sum_{n} \langle \psi^{*}_{n}\left(-{\bf k}\right) |S^{\dagger} \mathcal{I}^{\dagger} | \nabla_{\bf k}| \mathcal{I} S| \psi^{*}_{n}\left(-{\bf k}\right) \rangle \notag \\
&= i \sum_{mn} \langle \psi^{*}_{n}\left(-{\bf k}\right) |S^{\dagger} \mathcal{I}^{\dagger} | \nabla_{\bf k} |u_{m}\left(-{\bf k}\right)\rangle \langle u_{m}\left(-{\bf k}\right) | \mathcal{I} S| \psi^{*}_{n}\left(-{\bf k}\right) \rangle \notag \\
&= i \sum_{mn} \langle u_{m}\left(-{\bf k}\right) | \mathcal{I} S| \psi^{*}_{n}\left(-{\bf k}\right) \rangle \langle \psi^{*}_{n}\left(-{\bf k}\right) |S^{\dagger} \mathcal{I}^{\dagger} | \nabla_{\bf k} |u_{m}\left(-{\bf k}\right)\rangle \notag \\
&+ i \sum_{mn} \langle \psi^{*}_{n}\left(-{\bf k}\right) |S^{\dagger} \mathcal{I}^{\dagger} |u_{m}\left(-{\bf k}\right)\rangle  \nabla_{\bf k} \langle u_{m}\left(-{\bf k}\right) | \mathcal{I} S| \psi^{*}_{n}\left(-{\bf k}\right) \rangle \notag \\
&= - \mathcal{A}\left(-{\bf k}\right) +i\Tr \left[ M^{\dagger}\left({\bf k}\right) \nabla_{\bf k} M\left({\bf k}\right) \right] \notag \\
\mathcal{A}\left({\bf k}\right)&= - \mathcal{A}\left(-{\bf k}\right) +i~ \nabla_{\bf k} \ln\Det M\left({\bf k}\right) \notag
\end{align}
where $M_{mn}\left({\bf k}\right)=\langle \psi_{m}\left(-{\bf k}\right)| \mathcal{I}S | \psi^{*}_{n}\left(-{\bf k}\right) \rangle$ is a unitary matrix.

Moreover, the Berry connections for empty and occupied bands are related in the following way:
\begin{align}
\mathcal{A}^{+}\left({\bf k}\right) &= i \sum_{E_{n}>0} \langle \psi_{n}\left({\bf k}\right) | \nabla_{\bf k} | \psi_{n}\left({\bf k}\right) \rangle \notag \\
&= i \sum_{E_{n}<0} \langle \psi^{*}_{n}\left(-{\bf k}\right) | S^{\dagger} \nabla_{\bf k} S | \psi^{*}_{n}\left(-{\bf k}\right) \rangle \notag \\
&= i \sum_{E_{n}<0} \langle \psi^{*}_{n}\left(-{\bf k}\right) | \nabla_{\bf k} | \psi^{*}_{n}\left(-{\bf k}\right) \rangle \notag \\
&= i \sum_{E_{n}<0} \langle \psi_{n}\left(-{\bf k}\right) | \nabla_{-\bf k} | \psi_{n}\left(-{\bf k}\right) \rangle \notag \\
&= \mathcal{A}^{-}\left(-{\bf k}\right)~, \notag
\end{align}
where ``$\pm$'' respectively refers to the empty and occupied energy bands.

This results in
\begin{align}
\mathcal{A}\left({\bf k}\right) &= \frac{i}{2} \nabla_{\bf k} \ln\Det M\left({\bf k}\right) ~, \notag
\end{align}
which requires the Berry curvature to be null in the whole BZ: $\mathcal{F}\left({\bf k}\right)=\nabla_{\bf k}\times\mathcal{A}\left({\bf k}\right)=0$. It additionally verifies $\mathcal{F}^{\pm}\left({\bf k}\right)=\nabla_{\bf k}\times\mathcal{A^{\pm}}\left({\bf k}\right)=\mathcal{F}^{\pm}\left(-{\bf k}\right)$.

\subsubsection{Berry phase and Chern number}

The Bloch Hamiltonian matrix $\tilde H$ belongs to the Bogoliubov-de Gennes class D and the topology of its two-dimensional band structure is described by a first Chern number \cite{PhysRevB.78.195125}. Based on the negative-energy bands, the Chern number is defined by
\begin{align}\label{Appendix Chern Definition}
\nu &= \frac{1}{2\pi} \int_{BZ} d^{2}k ~ \mathcal{F}^{-}\left({\bf k}\right) \notag \\
&= \frac{1}{\pi} \int_{\frac{BZ}{2}} d^{2}k ~ \mathcal{F}^{-}\left({\bf k}\right)
\end{align}
Besides, the Berry phase along a path that encloses half the Brillouin zone $\frac{BZ}{2}$ is given by
\begin{align}\label{Appendix Berry Phase Definition}
\gamma_{\frac{BZ}{2}}&= \oint_{\frac{BZ}{2}} d{\bf k} \cdot \mathcal{A}^{-}\left({\bf k}\right) \\
&= \int_{\Gamma_{0}}^{\Gamma_{1}} d{\bf k} \cdot \mathcal{A}\left({\bf k}\right) + \frac{1}{\pi} \int_{\Gamma_{2}}^{\Gamma_{3}} d{\bf k} \cdot \mathcal{A}\left({\bf k}\right) \notag \\
&= i \ln \sqrt{\frac{\Det M\left({\bf \Gamma_{1}}\right)}{\Det M\left({\bf \Gamma_{0}}\right)}\frac{\Det M\left({\bf \Gamma_{3}}\right)}{\Det M\left({\bf \Gamma_{2}}\right)}} ~, \notag
\end{align}
Using the fact that, at the time-reversal invariant points,
\begin{align}
M_{mn}\left({\bf \Gamma_{i}}\right) &= \langle \psi_{m}\left(-{\bf \Gamma_{i}}\right)| \mathcal{I}S | \psi^{*}_{n}\left(-{\bf \Gamma_{i}}\right) \rangle \notag \\
&= \langle \psi_{m}\left({\bf \Gamma_{i}}\right)| \mathcal{I}S | \psi^{*}_{n}\left(-{\bf \Gamma_{i}}\right) \rangle \notag \\
&= \pi_{m}\left({\bf \Gamma_{i}}\right) \langle \psi_{m}\left({\bf \Gamma_{i}}\right)| S | \psi^{*}_{n}\left(-{\bf \Gamma_{i}}\right) \rangle \notag
\end{align}
and that $|\psi_{m}\left({\bf \Gamma_{i}}\right)\rangle$ and $S | \psi^{*}_{m}\left(-{\bf \Gamma_{i}}\right) \rangle$ are orthogonal eigenstates, one ends up with the following relation
\begin{align}
e^{i\gamma_{\frac{BZ}{2}}} &= \prod_{i=0}^{3}\sqrt{\prod_{n}\pi_{m}\left({\bf \Gamma_{i}}\right)} \notag \\
&= \prod_{i=0}^{3}\prod_{\text{occupied } n}\pi_{m}\left({\bf \Gamma_{i}}\right) \notag \\
&= \prod_{i=0}^{3}\delta_{i} \notag \\
&= \delta_{0}\delta_{2}
\end{align}

\begin{itemize}
\item If the spectrum is not gapped, as it may be the case for linear polarizations, then $\prod_{i=0}^{3}\delta_{i}=-1$ implies that the Berry phase satisfies $\gamma_{\frac{BZ}{2}}=\pi~[2\pi]$, and that there are Fermi points within $\frac{BZ}{2}$, so that the system lies in a semimetallic phase.

\item If the spectrum is gapped, as it is the case for elliptic polarizations, then Stokes theorem provides a relation between Eq.~\ref{Appendix Chern Definition} and Eq.~\ref{Appendix Berry Phase Definition}, which results in $e^{i\nu\pi}= \delta_{0}\delta_{2}$. When this quantity is negative, the Chern number is necessarily non-zero, and the system lies in a topological insulating phase.
\end{itemize}

\end{document}